# Phononic four-wave mixing

Authors: Adarsh Ganesan[1], Cuong Do[1], Ashwin Seshia[1]

1. Nanoscience Centre, University of Cambridge, Cambridge, UK

**We present the first experimental observations of phononic four-wave mixing (FWM) in a piezoelectrically actuated free-free beam microstructure. The FWM response is facilitated by the intrinsic coupling between a driven mode and an auto-parametrically excited sub-harmonic mode. Motivated by the experimental results, a dynamical model for FWM has been specified.**

Four-Wave Mixing (FWM) phenomenon is a nonlinear parametric process in which four frequency components are involved either as an input or output. In optics, FWM arises in a third-order nonlinear medium [1,2] and also leads to many applications including engineering of optical frequency comb [3,4]. The FWM phenomenon has also been observed in electrical systems and has been referred to as 'Inter-modulation distortion' [5]. Further, the FWM is also existent in nonlinear atom optics [6-8]. This paper, for the first time, presents the experimental observations of FWM in phononic domain. The phononic FWM is mediated by the interactions between drive tones and two phonon modes $Q_1$ and $Q_2$ and is modelled by the coupled dynamics.

$$\ddot{Q}_1 = -\omega_1^2 Q_1 - 2\zeta_1\omega_1\dot{Q}_1 + f_{d1}\cos(\omega_{d1}t) + f_{d2}\cos(\omega_{d2}t) + \alpha_{11}Q_1^2 + \alpha_{22}Q_2^2 + \beta_{111}Q_1^3 \quad (1\text{-}1)$$
$$+ \beta_{122}Q_1 Q_2^2$$

$$\ddot{Q}_2 = -\omega_2^2 Q_2 - 2\zeta_2\omega_2\dot{Q}_2 + \alpha_{12}Q_1 Q_2 + \beta_{112}Q_1^2 Q_2 \quad (1\text{-}2)$$

where $f_{d1}$ and $f_{d2}$ are drive levels of tones $\omega_{d1}$ and $\omega_{d2}$, $\alpha_{ij}$ & $\beta_{ijk}$ are quadratic and cubic coupling coefficients and $\zeta_{i=1,2}$ are damping coefficients. When the frequency of external driving $\omega_{d1}$ matches the resonant mode frequency $\omega_1$ and for sufficiently large amplitude values of drive $f_{d1}$, the resonant mode $\omega_{d1}$ and auto-parametrically triggered sub-harmonic mode $\omega_{d1}/2$ are excited. Through the quadratic and cubic coupling terms defined in the equation (1), the spectral lines of $cos(\omega_{d1} + p(\omega_{d2} - \omega_{d1})); p\,\epsilon\,Z$ are generated (See supplementary section S1 for the analysis). Similarly, if the drive power level $f_{d2}$ is sufficiently high, the auto-parametric excitation of the mode $Q_2$ occurs at $\omega_{d2}/2$ and hence the spectral lines of $cos(\omega_{d2} + p(\omega_{d2} - \omega_{d1})); p\,\epsilon\,Z$ are generated.

The FWM mechanism is experimentally demonstrated using an extensional resonance mode (~3.86 MHz) in a microscopic free-free beam structure of dimensions 1100 x 350 x 11 µm (Figure 1A). A 0.5 µm thick AlN layer is deposited on the silicon device layer for piezoelectric transduction of the response (See supplementary section S2). The electrical response and out-of-plane (OOP) mechanical displacement profile of the resonator are obtained through spectrum analyser (SA) and Laser Doppler Vibrometry (LDV) measurements respectively. The in-plane (IP) and OOP displacements of the 3.86 MHz mode are linearly related and hence, the perturbation of the in-plane response is studied through the changes in OOP displacement obtained through vibrometry (Supplementary section S3). The drive tones are applied to the resonator through the split electrodes on the resonator. At a particular drive condition $S_{in}(\omega_{d1}) = 4\ dBm;\ S_{in}(\omega_{d2}) = 12\ dBm$, spectral lines are generated at $(1+n)\omega_{d2} - n\omega_{d1};\ n \in Z$ (Figure 1B). Also, Laser Doppler Vibrometry (LDV) measurements reveal the distinctive role of an auto-parametrically generated sub-harmonic resonant mode [9] (Figures 1B and 1C) in FWM. Figure 1C shows the presence of additional spectral lines (Figure 1c2) at the antinodes of the sub-harmonic resonant mode in contrast to the nodes (Figure 1c1).

To understand the dependence of drive conditions on the FWM process, drive levels $S_{in}(\omega_{d1} = 3.855\ MHz)$ and $S_{in}(\omega_{d2} = 3.86\ MHz)$ are varied from $0\ dBm$ to $20\ dBm$. The raw data set is provided in the supplementary section S4. In this data set, there exist four distinct types of spectral responses namely 'Nominal', 'FWM-1', 'FWM-2', 'High-order FWM'. In the 'Nominal regime', FWM does not occur and only the spectral lines corresponding to the drive tones are present (Figure 2a1). This is attributed to the absence of auto-parametric excitation. However, once the threshold for auto-parametric excitation is met, the FWM mechanism is established and additional spectral lines are consequently produced (Figures 2a2 and 2a3). Specifically, in 'FWM-1', the resonant dynamics of drive tone ($\omega_{d1}$) gets periodically modulated by the drive tone ($\omega_{d2}$) and hence, the additional spectral lines are produced on either sides of $\omega_{d1}$ (Figure 2a3). In contrast, FWM-2 results in the spectral lines about $\omega_{d2}$ (Figure 2a2). In addition, there is an evidence for high-order FWM regime at higher drive levels of both $\omega_{d1}$ and $\omega_{d2}$ (Figure 2a4). Here, extra tones are produced and the net spectral line spacing halves. This behaviour is qualitatively similar to the nonlinear process corresponding to the spectral line growth previously seen in optical microresonators [10]. However, in phononic FWM, the excitation of these interleaved spectral lines can be attributed to an additional nonlinear interplay between FWM processes mediated by both the drive tones.

The drive conditions are thus relevant in determining the solutions of FWM. The spatial effects on phonon modes due to FWM have interesting features of note. Firstly, in the absence of FWM, only the spectral lines corresponding to the drive tones are observed. 'Laser Doppler Vibrometry (LDV)' images are recorded at these frequencies and the concomitant RMS vibration profile is obtained. These correspond to the driven resonant modeshape. However, in the FWM regime, the mechanical vibrations are also observed at the frequencies $(\omega_{d2} + n(\omega_{d2} - \omega_{d1}));\ n = \pm 1, \pm 2$ (Figures 2b1-

2b6). The LDV data revealed the dependence of FWM on sub-harmonic mode excitation. At the nodal regions of subharmonic resonant mode, phonons of frequencies $\omega_{d1}$ and $\omega_{d2}$ are only present. In contrary, at the locations corresponding to the antinodes of sub-harmonic resonant mode, phonons of frequencies $\omega_{d1}$, $\omega_{d2}$ and $2\omega_{d2} - \omega_{d1}$ are generated. This feature can also be noted by comparing vibration patterns at $\omega_{d1}$ and $2\omega_{d2} - \omega_{d1}$ to the sub-harmonic resonant modeshape. Ideally, at $\omega_{d1}$, there should be an overlap of driven and parametrically excited sub-harmonic modeshapes. However, due to insignificant direct excitation at $\omega_{d1}$, the phonons of frequency $\omega_{d1}$ are mostly associated at the antinodes of sub-harmonic modes corresponding to the escalated excitation of phonons through FWM (Figure 2b3). In contrary, at $\omega_{d2}$, there is also significant direct excitation and hence, the merger of driven and parametrically excited sub-harmonic resonant modeshape is observed (Figure 2b4). At the other spectral locations $\left(\omega_{d2} + n(\omega_{d2} - \omega_{d1})\right); n \in Z - \{-1\}$, phonons are continuously generated at the antinodes of sub-harmonic mode. At the highest displacement location of sub-harmonic component, higher order expansions are involved and far-enough spectral lines are generated. In contrast, at the lowest displacement of sub-harmonic component, the phonons of frequency $\left(\omega_{d2} + n(\omega_{d2} - \omega_{d1})\right); n \in Z - \{-1\}$ are absent. This behaviour results in a continuous decay in the depth of sub-harmonic modeshapes at far-enough separations from the frequency $\omega_{d2}$ (Figures 2b1, 2b2, 2b5 and 2b6). The RMS vibration pattern (Figure 2b7) corresponding to the frequencies $\omega_{d2}$ and $\left(\omega_{d2} + n(\omega_{d2} - \omega_{d1})\right); n \in Z$ reveals the merger of the driven and parametrically excited resonant modeshapes. When the FWM mechanism is prominent i.e. at high drive levels, the phonon population at the antinodes of parametrically excited mode is significantly higher than at the antinodes of driven mode. Hence, as the signal level $S_{in}(\omega_{d2})$ is increased, the RMS vibration pattern continuously shifts from the driven phonon mode shape to the sub-harmonic mode shape after crossing the parametric excitation threshold $10\ dBm$ (Figures 2c1-2c7). In FWM-1, the spatial re-distribution occurs at drive tone $\omega_{d2}$ concomitant to the reversal of FWM process between the drive tones (Figures 2d1-2d6).

The generation of additional spectral lines through higher order expansions is shown in the figure 3A. Here, the drive signal $S_{in}(\omega_{d2} = 3.86\ MHz)$ is kept constant at $14\ dBm$ and $S_{in}(\omega_{d1} = 3.855\ MHz)$ is increased from $0\ dBm$ to $22\ dBm$. At far enough drive levels of $S_{in}(\omega_{d1} = 3.855\ MHz) \sim < 8\ dBm$, the FWM mechanism is absent and the two drive tones are only observed. As mentioned before, once the threshold is met, the spectral lines $\omega_{d2} \pm (\omega_{d2} - \omega_{d1}) \equiv 2\omega_{d2} - \omega_{d1}\ \&\ \omega_{d1}$ are generated. When the drive level of $S_{in}(\omega_{d1} = 3.855\ MHz)$ is further increased, higher order ($\kappa$) expansions take place and consequently additional spectral lines $\omega_{d2} \pm \kappa(\omega_{d2} - \omega_{d1})$ are produced. The steps in figure 3A indicate the inherent thresholds corresponding to the higher order expansions. To also test the relevance of the spacing between the frequency of drive tones and resultant spectral line spacing $g$, the drive frequency $\omega_{d1}$ is varied. As seen in the figure 3B, the spacing $g$ equals $|\omega_{d1} - \omega_{d2}|$. The frequency responses corresponding to different drive frequencies $\omega_{d1}$ are provided in figure 3C. At close-enough separation between $\omega_{d1}$ and $\omega_{d2}$, the spectral lines are less well-defined (Figures 3c4, 3c5, 3c7 and 3c8). Here, multiple spectral lines are formed through high-order FWM (Figures 3c4 and 3c8) and ultimately results in the filling-in of phonon mode dispersion (Figures 3c5 and 3c7).

In summary, this paper experimentally demonstrates the mechanism of phononic Four-Wave Mixing. The potential applications of phononic FWM include phonon lasing [11], phonon computation [12-13] and nonlinear phononics [14].

**Acknowledgements**

Funding from the Cambridge Trusts is gratefully acknowledged.

**Authors' contributions**

AG and CD designed the device and performed the experiments; AG and AAS analyzed the results and wrote the manuscript; AAS supervised the research.

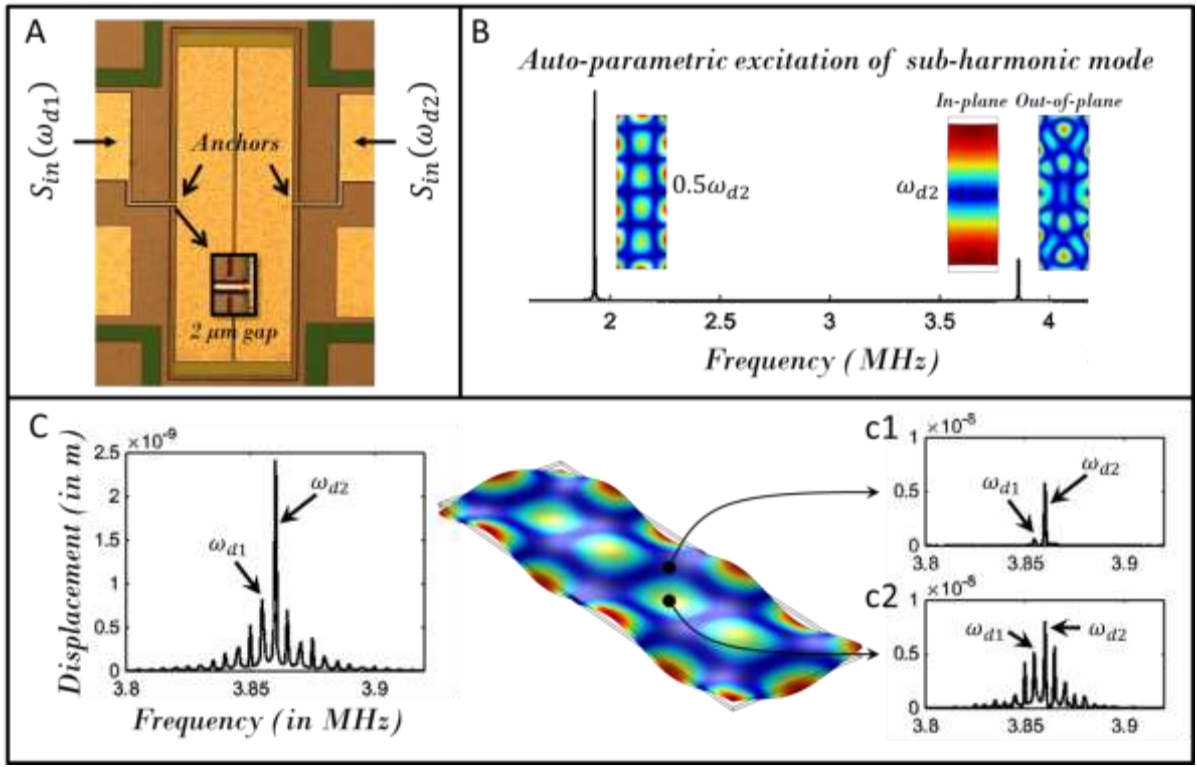

Figure 1: **Phononic Four-Wave Mixing.** A: Signals $S(\omega_{d1})$ and $S(\omega_{d2})$ are applied on a free-free beam microstructure; B: An auto-parametric excitation of sub-harmonic mode (out-of-plane) at high drive power levels (> 8dBm) of in-plane extensional mode with detectable out-of-plane component; C: Left: Surface average displacement spectrum at the $S(\omega_{d1} = 3.855\ MHz) = 4\ dBm$ and $S(\omega_{d2} = 3.86\ MHz) = 12\ dBm$ demonstrating FWM; Middle: The sub-harmonic mode shape; Right-top- c1: The displacement at the node of sub-harmonic mode indicating the absence of FWM; Right-down- c2: The displacement at the antinode of sub-harmonic mode indicating the presence of FWM.

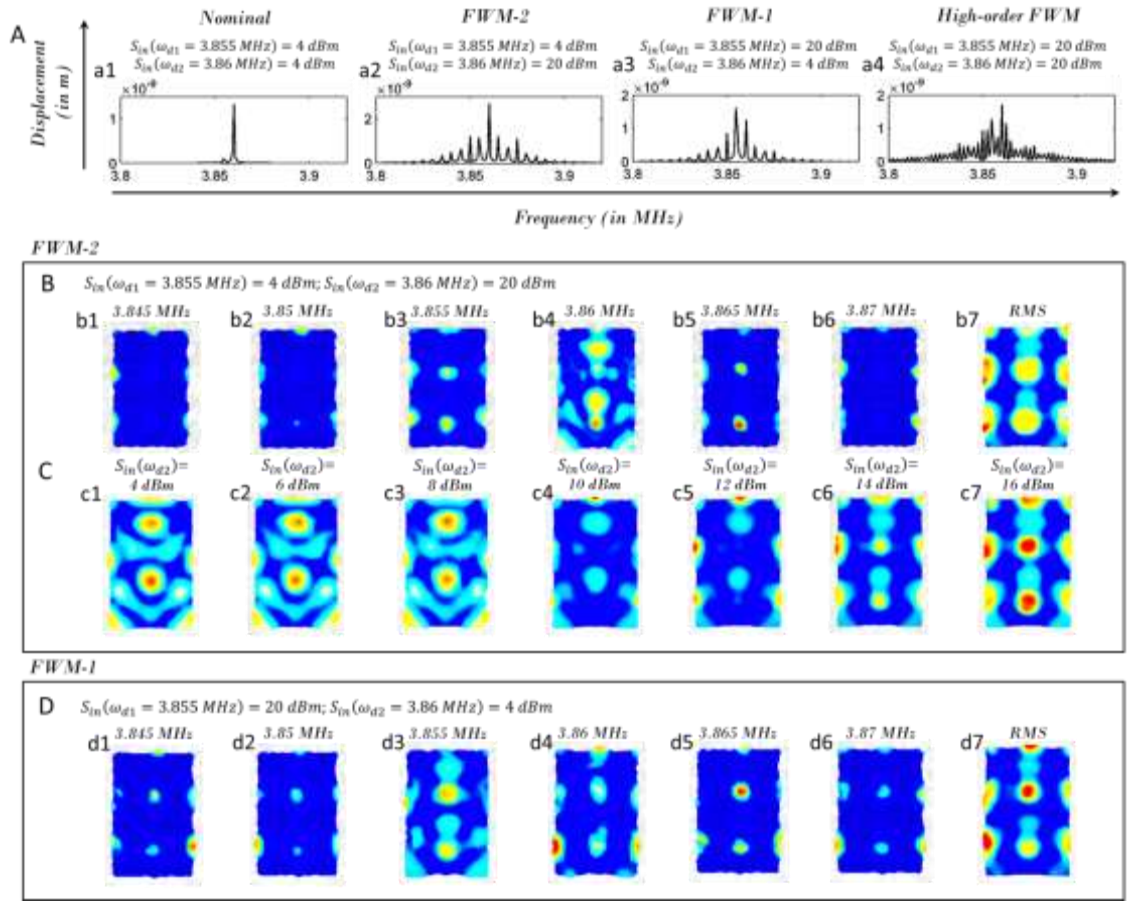

Figure 2: **Regime of Phononic Four-Wave Mixing.** A: Representative frequency responses corresponding to a1: Nominal regime: $S_{in}(\omega_{d1} = 3.855\ MHz) = 4\ dBm$ and $S_{in}(\omega_{d2} = 3.86\ MHz) = 4\ dBm$; a2: FWM-2 regime: $S_{in}(\omega_{d1} = 3.855\ MHz) = 4\ dBm$ and $S_{in}(\omega_{d2} = 3.86\ MHz) = 20\ dBm$; a3: FWM-3 regime: $S_{in}(\omega_{d1} = 3.855\ MHz) = 20\ dBm$ and $S_{in}(\omega_{d2} = 3.86\ MHz) = 4\ dBm$; a4: High-order FWM regime: $S_{in}(\omega_{d1} = 3.855\ MHz) = 20\ dBm$ and $S_{in}(\omega_{d2} = 3.86\ MHz) = 20\ dBm$; B-D: Laser Doppler Vibrometry data images; b1-b6: Surface displacement profiles of the device at the frequencies 3.845, 3.85, 3.855, 3.86, 3.865 and 3.87 MHz and b7: RMS surface displacement profile for the frequency band 3.7-4 MHz corresponding to the drive conditions $S_{in}(\omega_{d1} = 3.855\ MHz) = 4\ dBm$ and $S_{in}(\omega_{d2} = 3.86\ MHz) = 20\ dBm$; c1-c7: RMS surface displacement profile for the frequency band 3.7-4 MHz corresponding to the drive conditions $S_{in}(\omega_{d1} = 3.855\ MHz) = 4, 6, 8, 10, 12, 14\ and\ 16\ dBm$ and $S_{in}(\omega_{d2} = 3.86\ MHz) = 4\ dBm$; d1-d6: Surface displacement profiles at the frequencies 3.845, 3.85, 3.855, 3.86, 3.865 and 3.87 MHz and d7: RMS surface displacement profile for the frequency band 3.7-4 MHz corresponding to the drive conditions $S_{in}(\omega_{d1} = 3.855\ MHz) = 20\ dBm$ and $S_{in}(\omega_{d2} = 3.86\ MHz) = 4\ dBm$

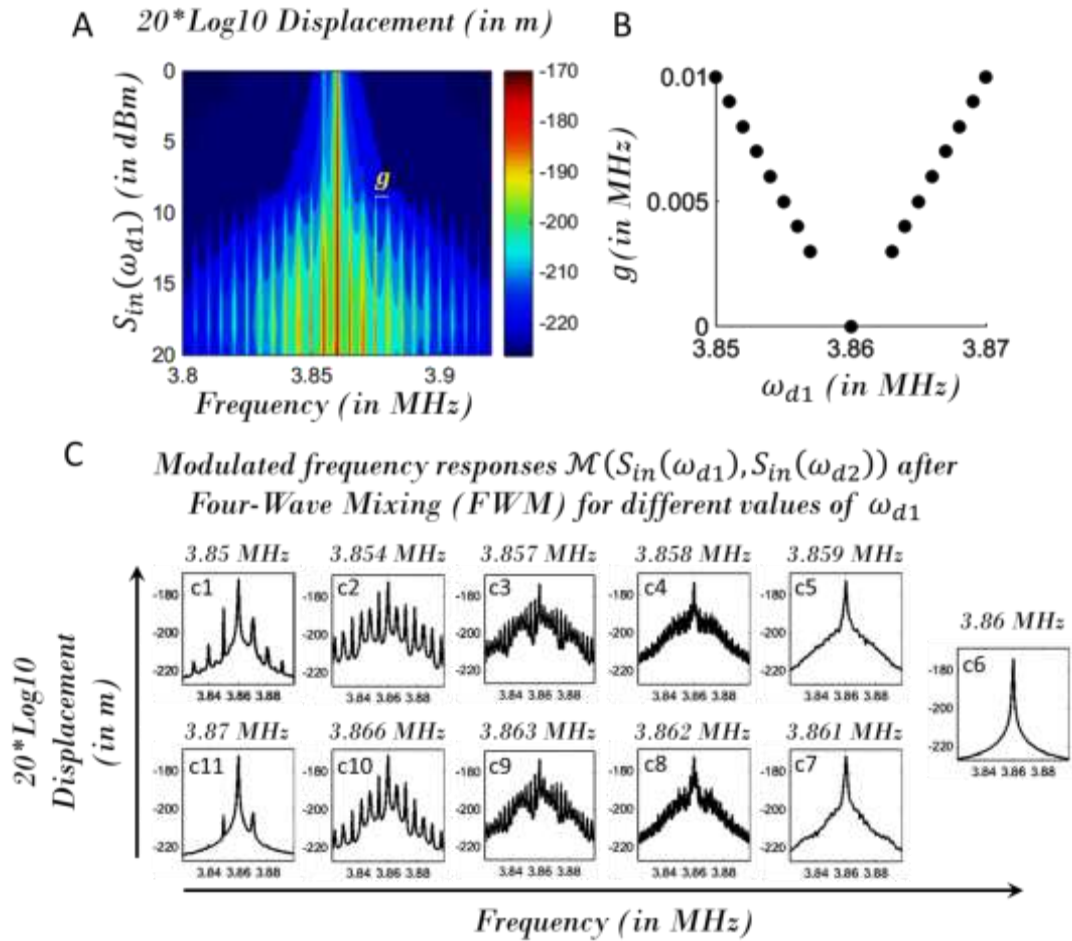

Figure 3: **Dependence of drive amplitude and frequency.** A: Frequency contour for different drive signal levels $S_{in}(\omega_{d1} = 3.855\ MHz)$; B: The spectral line spacing $g$ for different drive frequency $\omega_{d1}$ at a constant signal level $S_{in}(\omega_{d1}) = 4\ dBm$; C: The frequency responses corresponding to different drive frequencies $\omega_{d1} = 3.85, 3.854, 3.857\ 3.858, 3.859, 3.86, 3.861, 3.862, 3.863, 3.866$ and $3.87\ MHz$ (c1-c11) for $S_{in}(\omega_{d1}) = 4\ dB$. Note: The drive signal $S_{in}(\omega_{d2} = 3.86\ MHz) = 14\ dBm$ is kept constant during these measurements.

**Supplementary Information**

**Phononic four-wave mixing**

Authors: Adarsh Ganesan[1], Cuong Do[1], Ashwin Seshia[1]

1. Nanoscience Centre, University of Cambridge, Cambridge, UK

**1. Analytical derivation of the Phononic Four-Wave Mixing response**

For a truncated phase space spanned by two phonon modes $Q_1$ & $Q_2$, the coupled dynamics can be written as,

$$\ddot{Q}_1 = -\omega_1^2 Q_1 - 2\zeta_1\omega_1\dot{Q}_1 + f_{d1}\cos(\omega_{d1}t) + f_{d2}\cos(\omega_{d2}t) + \alpha_{11}Q_1^2 + \alpha_{22}Q_2^2 + \beta_{111}Q_1^3 \quad (1\text{-}1)$$
$$+ \beta_{122}Q_1 Q_2^2$$

$$\ddot{Q}_2 = -\omega_2^2 Q_2 - 2\zeta_2\omega_2\dot{Q}_2 + \alpha_{12}Q_1 Q_2 + \beta_{112}Q_1^2 Q_2 \quad (1\text{-}2)$$

where $f_{d1}$ and $f_{d2}$ are the displacements of drive tones $\omega_{d1}$ and $\omega_{d2}$ and $\alpha_{ij}$ & $\beta_{ijk}$ are quadratic and cubic coupling coefficients of FPU chain.

Let us consider the case $f_{d1} \gg f_{d2}$. The 0$^{th}$ order solution of $Q_1$ will be just a linear oscillator with the tone $cos(\omega_{d1}t)$. The $Q_1^{(0)}$ solution is now coupled into the dynamics of $Q_2$ to obtain $Q_2^{(0)}$. This case presents a Mathieu framework. The parametric loading offered by $Q_1$ on $Q_2$ triggers the mode at frequency $\omega_{d1}/2$. The inter-modulation among drive tones and sub-harmonic tone is established through quadratic and cubic nonlinear terms in (S1) and the tones $cos(\omega_{d1}t)$ and $cos((\omega_{d2} - \omega_{d1})t)$ are produced in the truncated frequency spectrum of $Q_1^{(1)}$. By solving equation (S1) through sequential iteration, the coupling between tones at $(\omega_{d2} - \omega_{d1})$ and $\omega_{d1}$ is established through the quadratic and cubic nonlinear terms in (S1) and the near-resonant terms of $cos(\omega_{d1} + p(\omega_{d2} - \omega_{d1})t); p \in Z$ are generated. In addition, the tones at $cos\left(\frac{\omega_{d1}}{2} + p(\omega_{d2} - \omega_{d1})t\right)$ and $cos(p(\omega_{d2} - \omega_{d1})t); p \in Z$ are generated.

**Supplementary section S2**

Operation of piezoelectrically driven micromechanical resonator

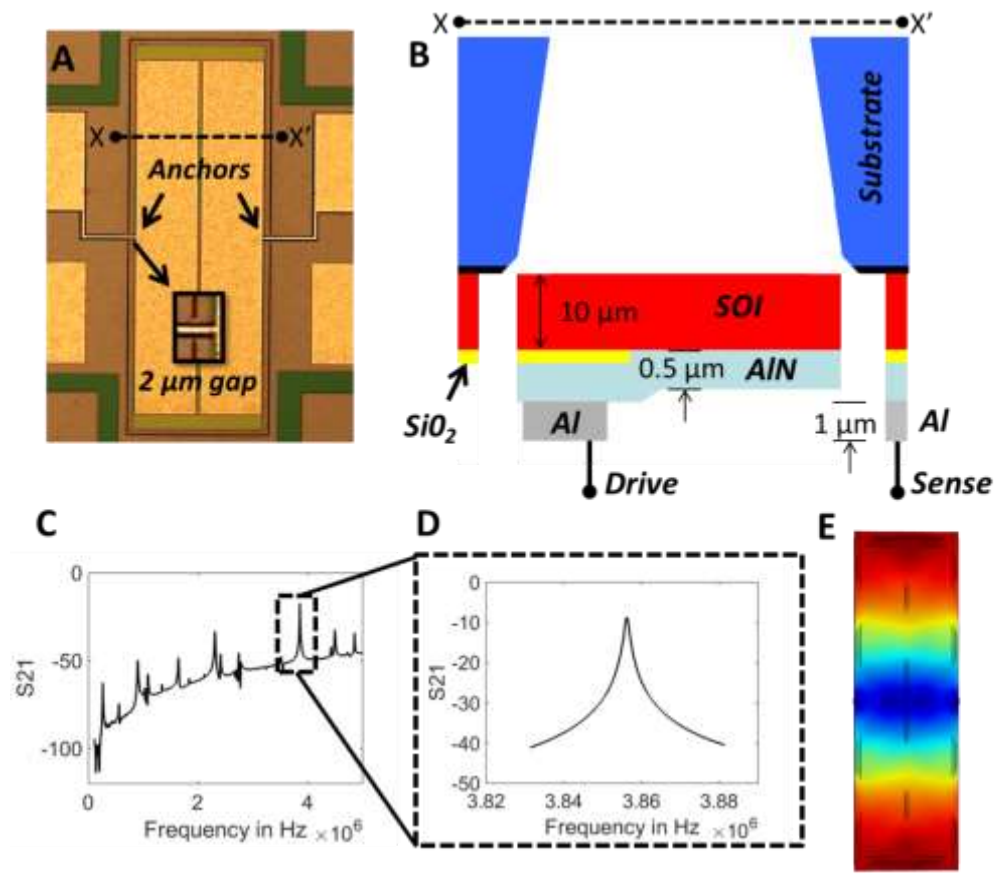

**Figure S1: Operation of piezoelectrically driven micromechanical resonator**: **A**: Free-free beam topology with 2 µm air gap for in-plane mode excitation; **B**: 1 µm thick Al electrodes patterned on 0.5 µm thick AlN piezoelectric film which is in-turn patterned on SOI substrate; the 10 µm thick SOI layer is then released through back-side etch to realize mechanical functionality; **C**: The scattering parameter S21 denoting forward transmission gain across a broad range of frequencies from 0-4.5 MHz; **D**: across a smaller range of frequencies 3.83-3.88 MHz; **E**: Resonant mode shape of ~3.86 MHz vibrations from eigen-frequency analysis in COMSOL and Doppler shift vibrometry.

**Supplementary section S3**

Laser Doppler Vibrometry measurement

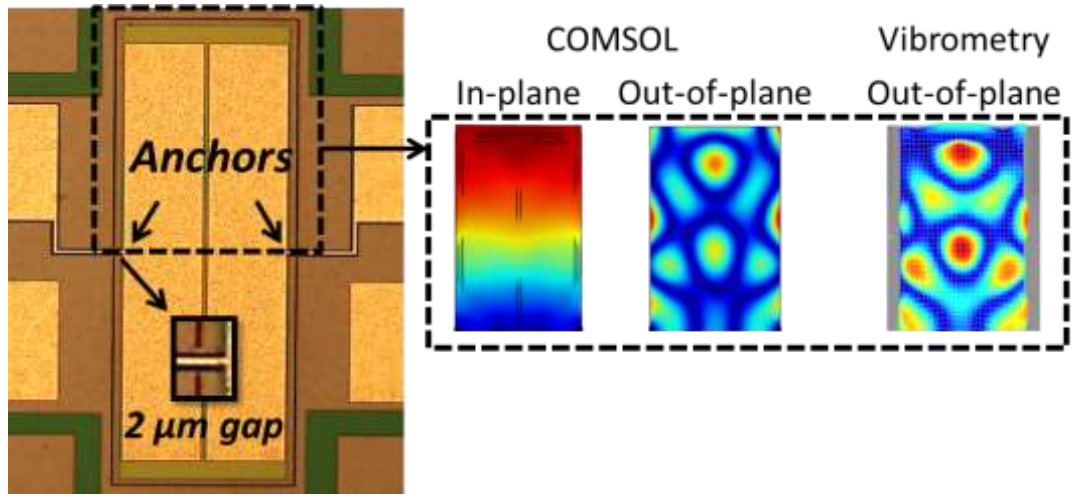

**Figure S2:** Laser Doppler Vibrometry is used to image the out-of-place surface displacement of piezoelectrically driven micro-mechanical resonator. The dotted square boundary represents the image capture area. The imaged surface displacement and the eigenfrequency analytical prediction based on COMSOL is presented.

## Supplementary section S4

Frequency responses and RMS surface displacement profiles obtained using Laser Doppler Vibrometry at different drive conditions

| $S_{in}(\omega_{d1})$ | $S_{in}(\omega_{d2})$ | Displacement (in m) vs. Frequency (in Hz) | RMS surface displacement profile |
|---|---|---|---|
| 0 | 0 | | |
| 0 | 2 | | |
| 0 | 4 | | |
| 0 | 6 | | |
| 0 | 8 | 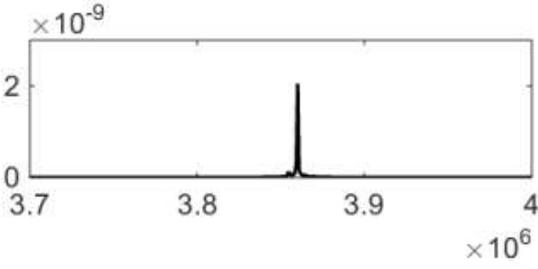 | 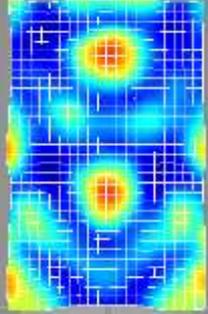 |
| 0 | 10 | 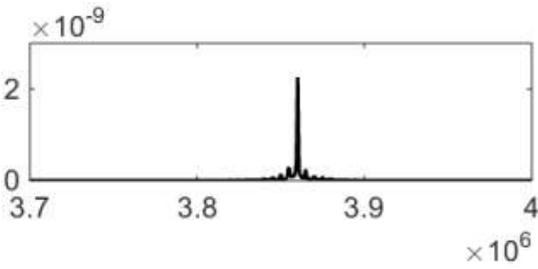 | 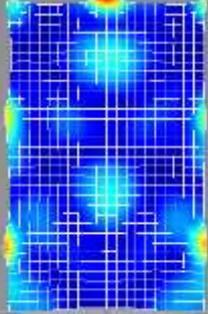 |
| 0 | 12 | 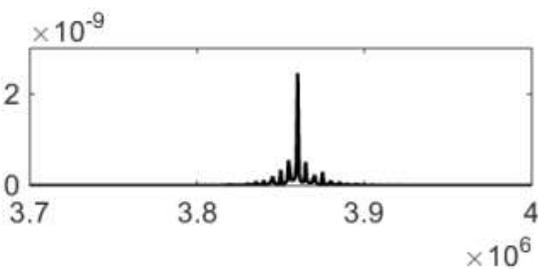 | 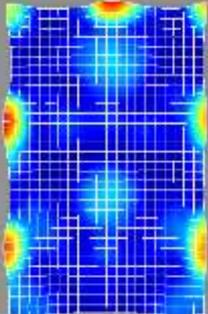 |
| 0 | 14 | 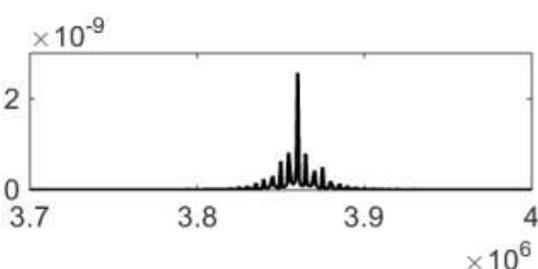 | 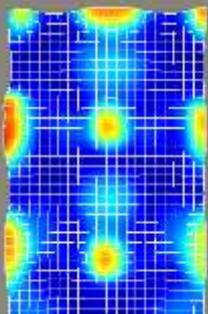 |
| 0 | 16 | 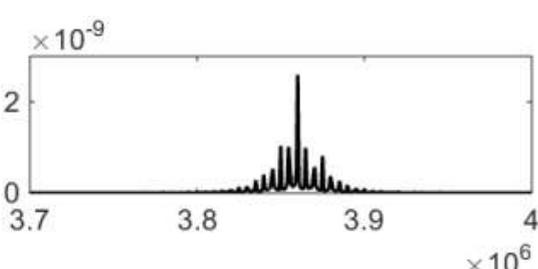 | 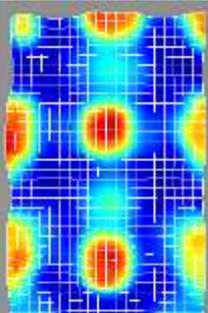 |
| 0 | 18 | 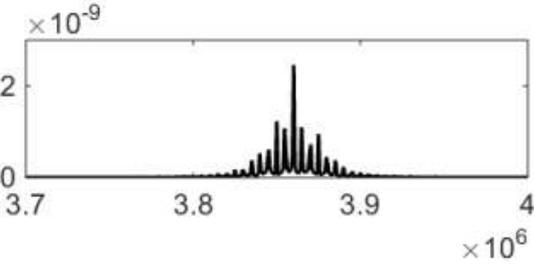 | 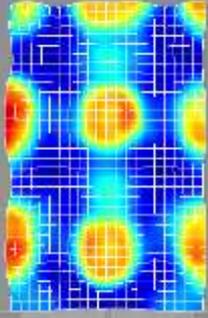 |
| 0 | 20 | 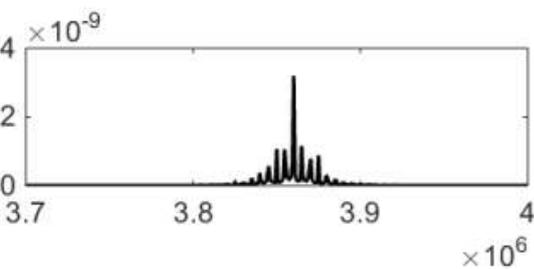 | 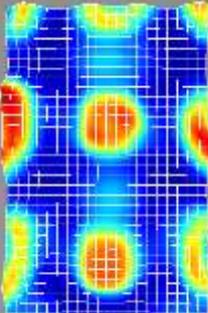 |
| 2 | 0 | 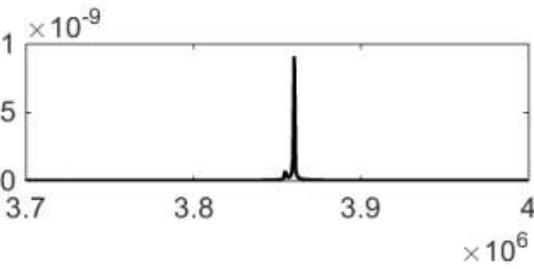 | 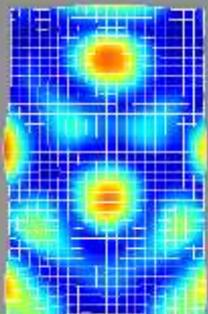 |
| 2 | 2 | 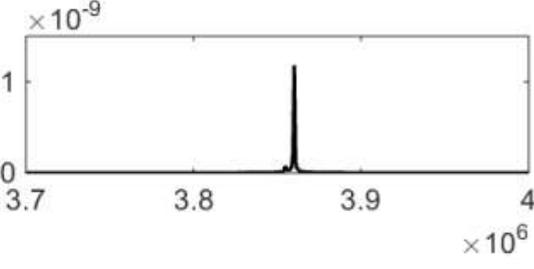 | 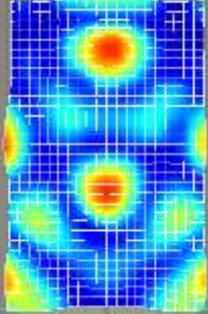 |
| 2 | 4 | 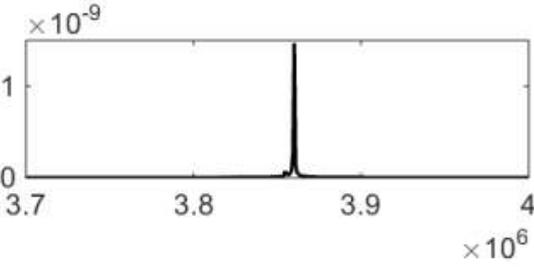 | 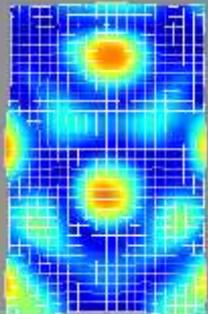 |
| 2 | 6 | 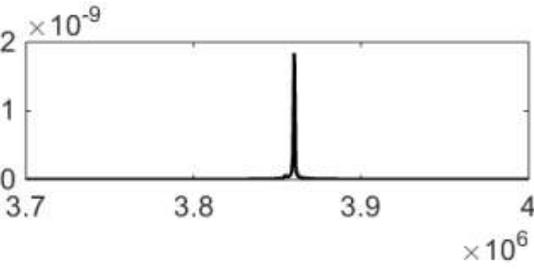 | 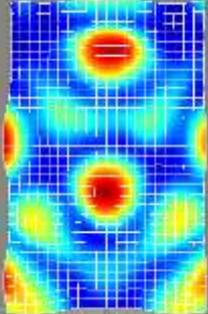 |
| 2 | 8 | 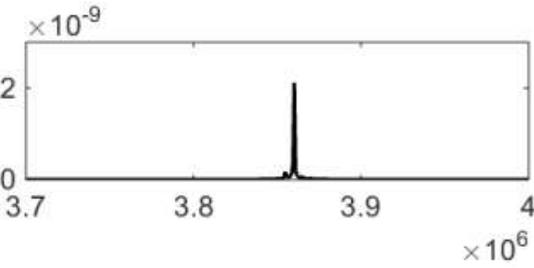 | 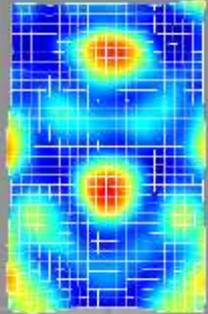 |
| 2 | 10 | 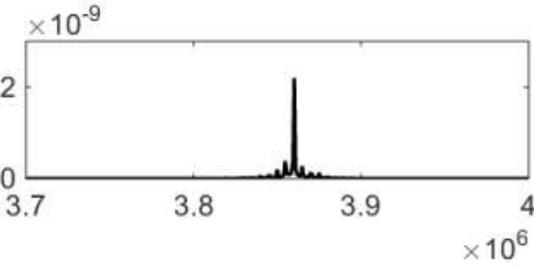 | 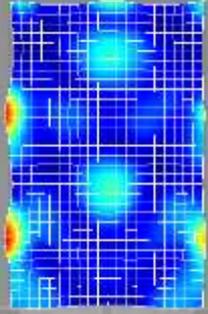 |
| 2 | 12 | 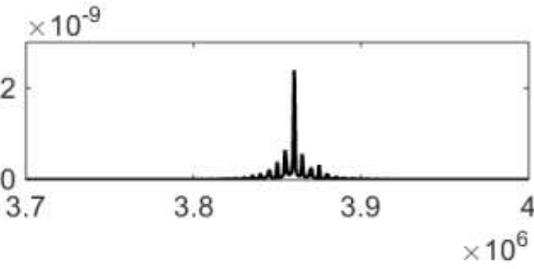 | 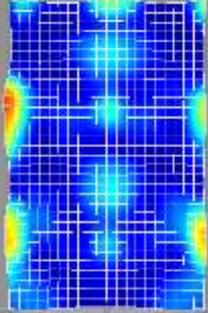 |
| 2 | 14 | 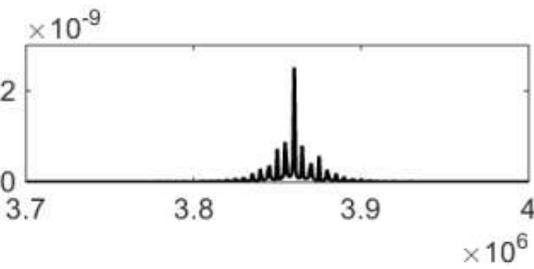 | 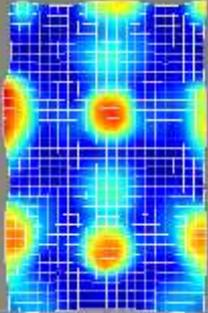 |
| 2 | 16 | 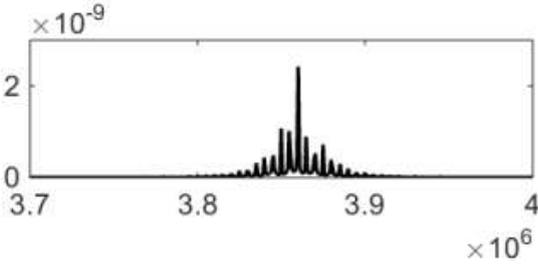 | 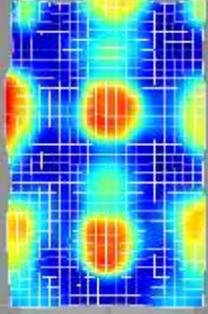 |
| 2 | 18 | 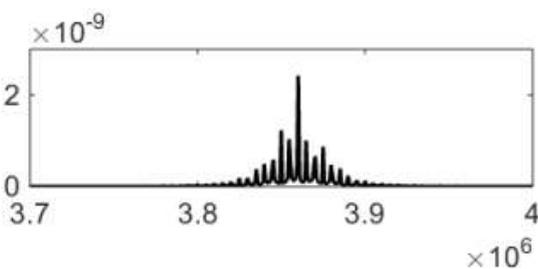 | 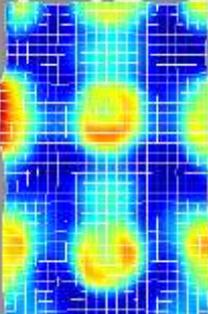 |
| 2 | 20 | 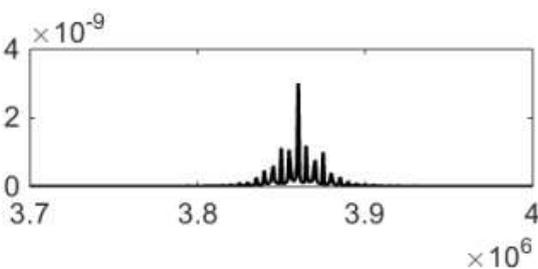 | 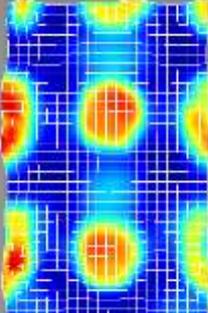 |
| 4 | 0 | 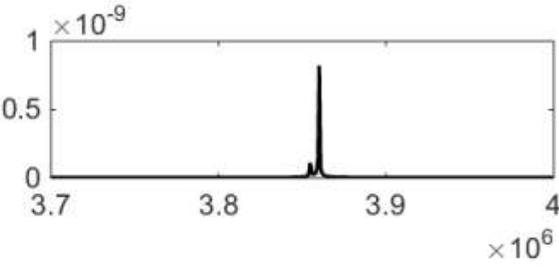 | 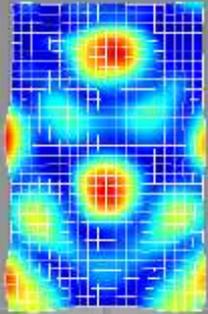 |
| 4 | 2 | 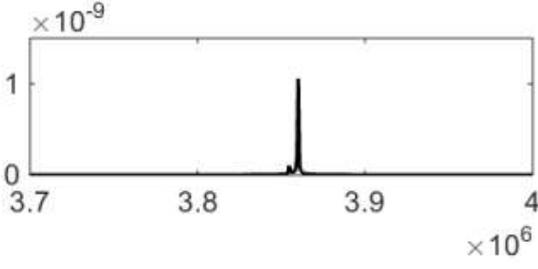 | 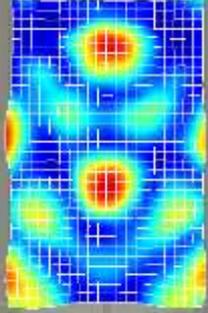 |
| 4 | 4 | 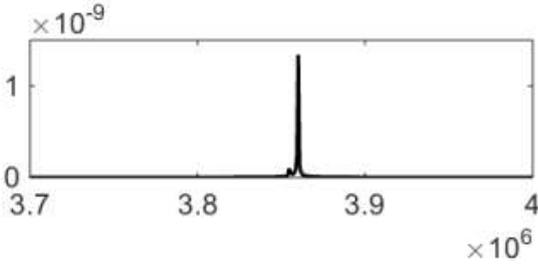 | 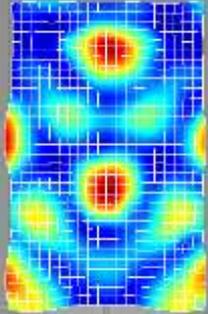 |
| 4 | 6 | 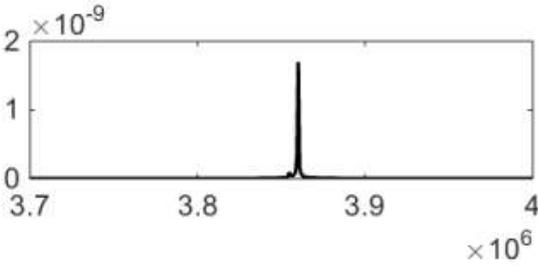 | 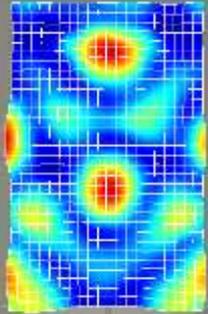 |
| 4 | 8 | 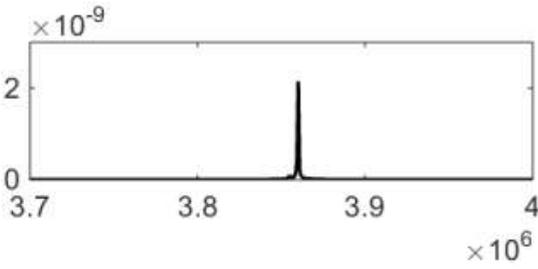 | 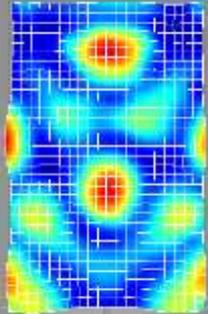 |
| 4 | 10 | 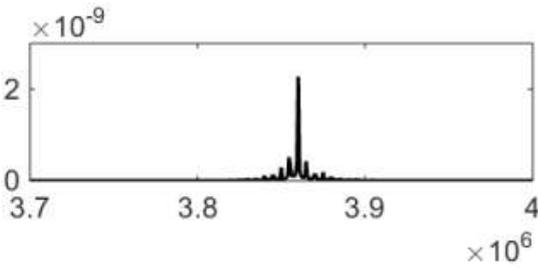 | 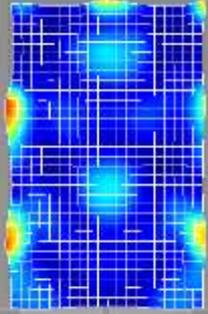 |
| 4 | 12 | 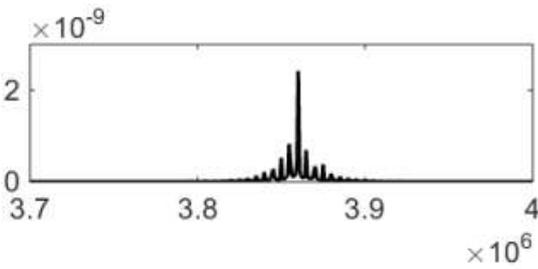 | 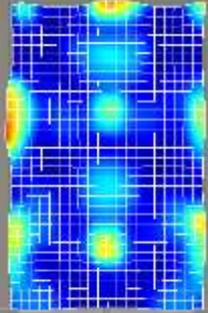 |
| 4 | 14 | 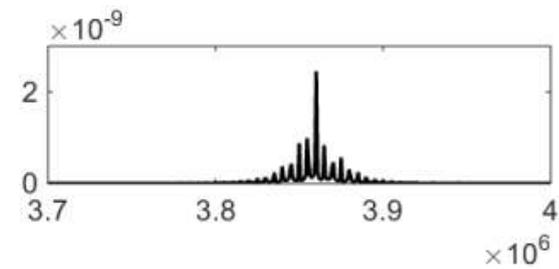 | 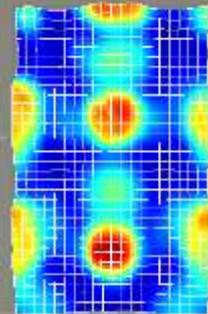 |
| 4 | 16 | 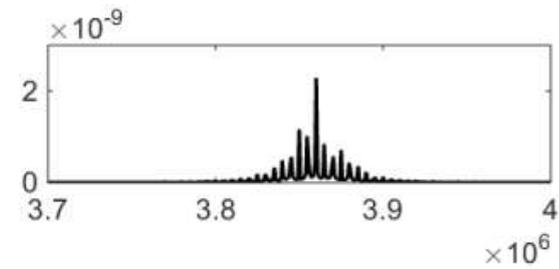 | 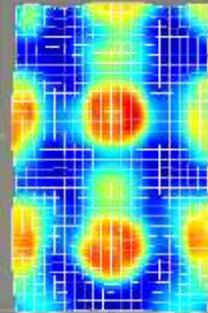 |
| 4 | 18 | 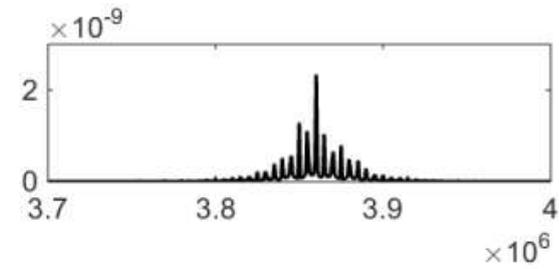 | 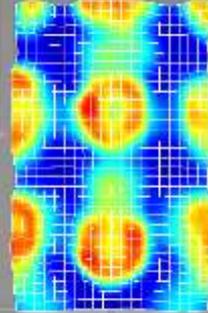 |
| 4 | 20 | 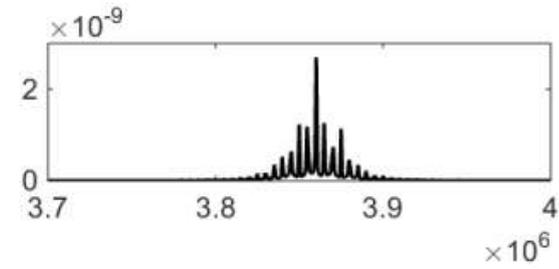 | 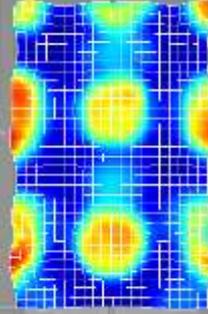 |
| 6 | 0 | 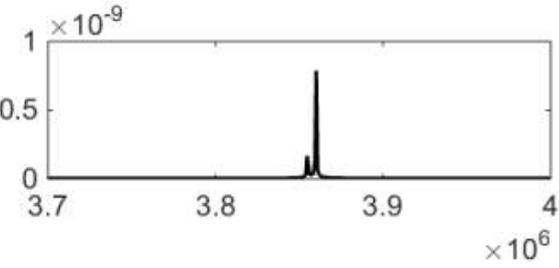 | 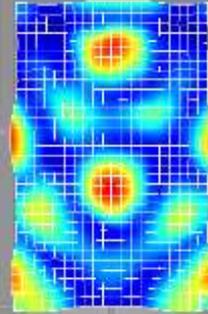 |
| 6 | 2 | 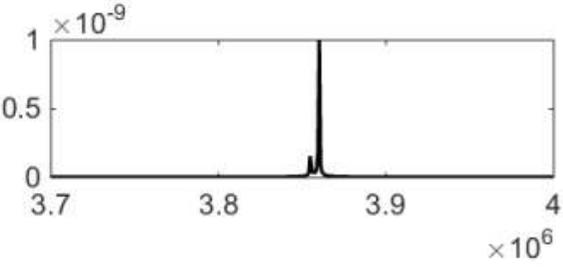 | 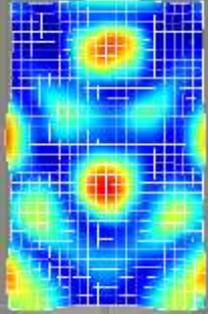 |
| 6 | 4 | 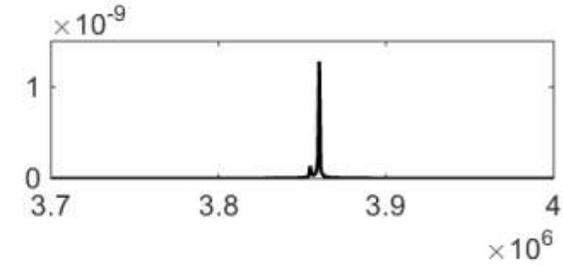 | 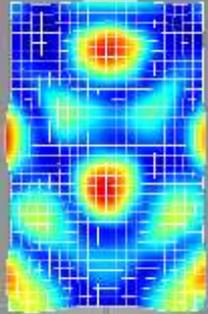 |
| 6 | 6 | 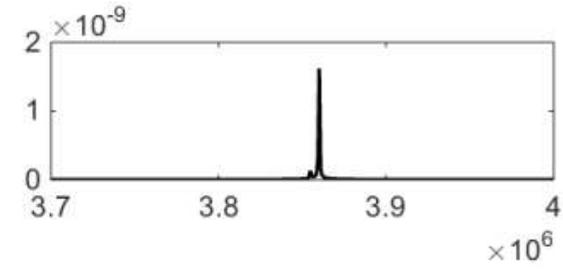 | 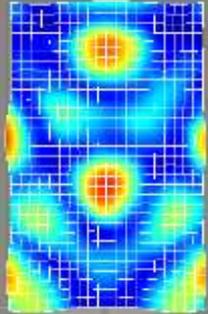 |
| 6 | 8 | 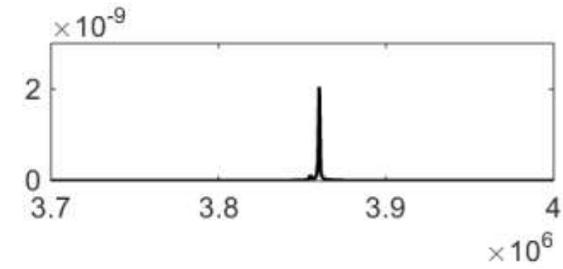 | 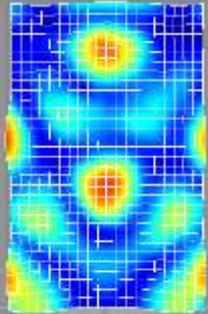 |
| 6 | 10 | 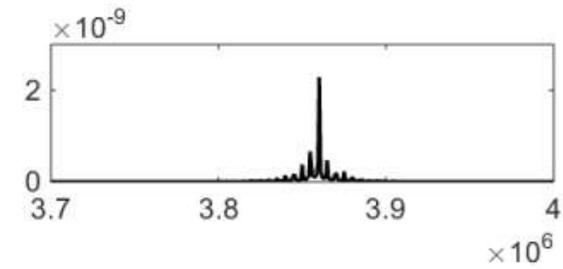 | 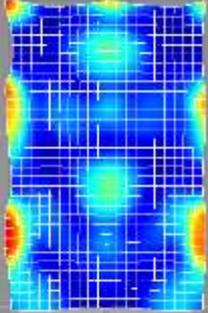 |
| 6 | 12 | 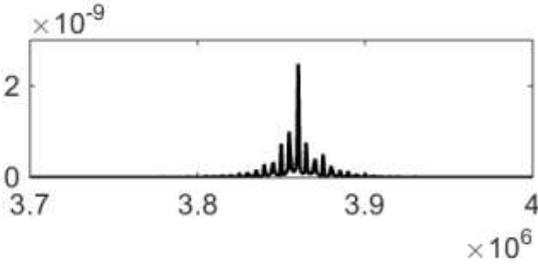 | 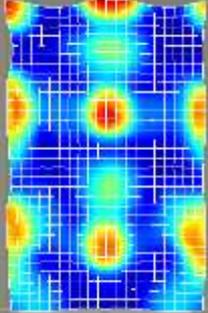 |
| 6 | 14 | 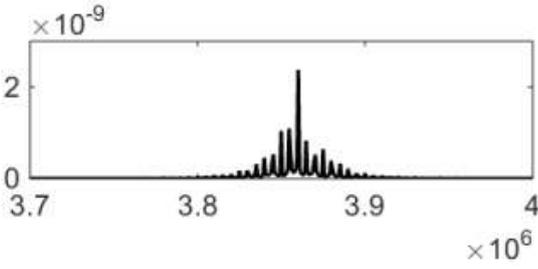 | 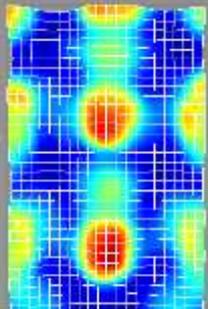 |
| 6 | 16 | 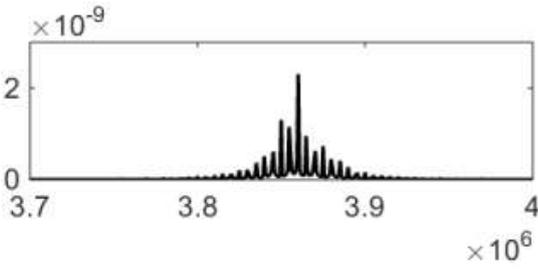 | 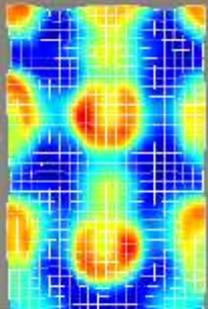 |
| 6 | 18 | 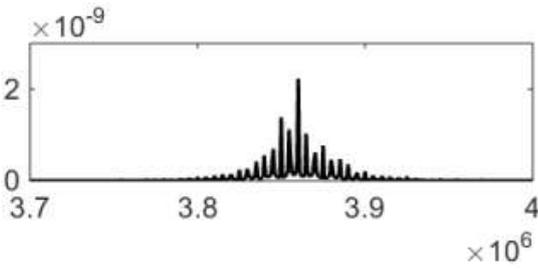 | 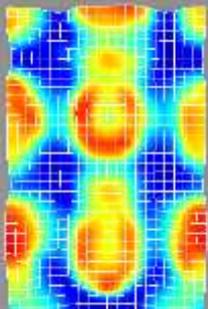 |
| 6 | 20 | 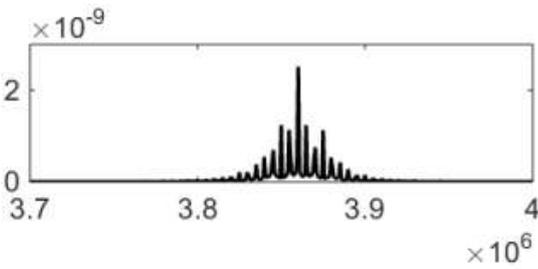 | 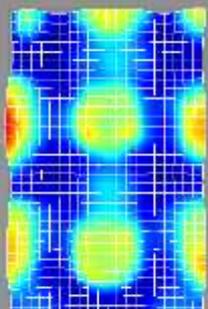 |
| 8 | 0 | 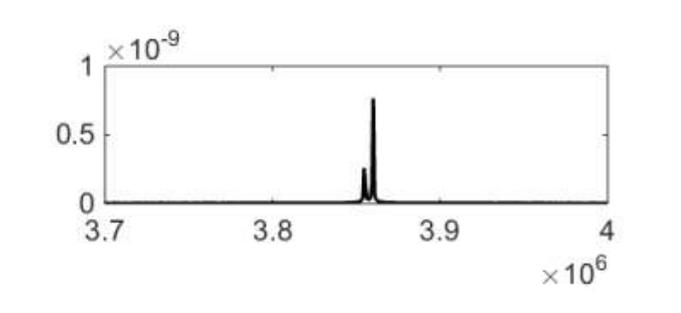 | 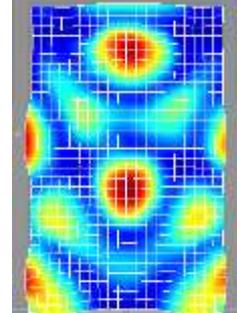 |
| 8 | 2 | 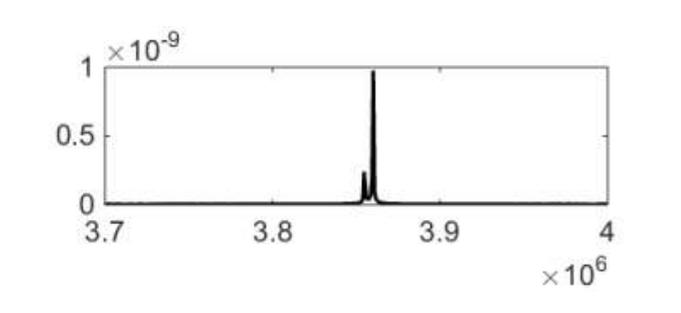 | 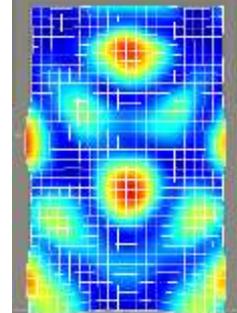 |
| 8 | 4 | 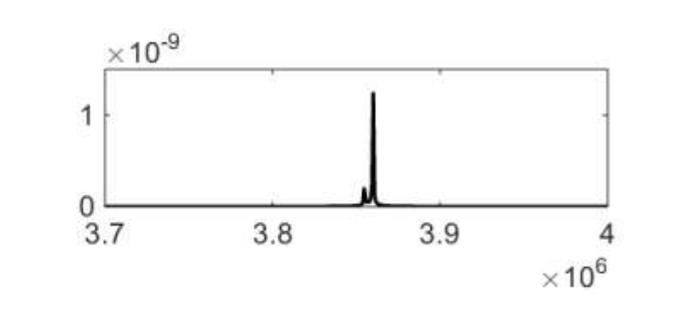 | 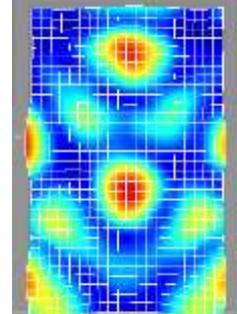 |
| 8 | 6 | 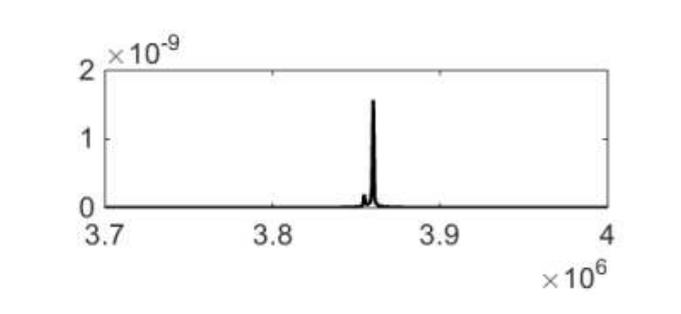 | 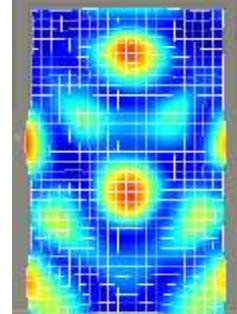 |
| 8 | 8 | 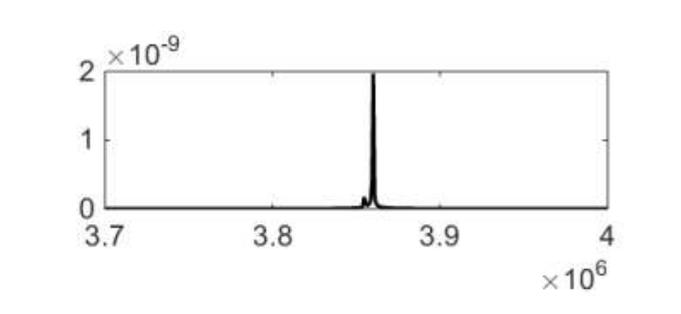 | 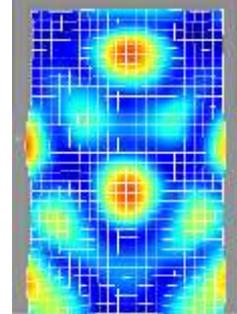 |
| 8 | 10 | 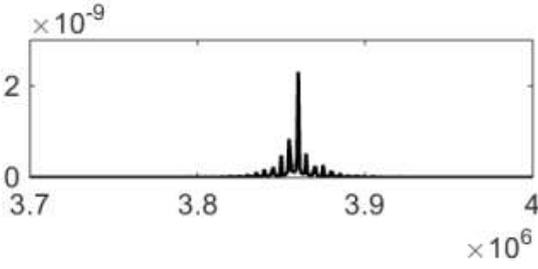 | 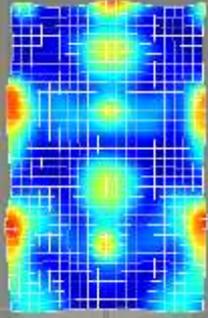 |
| 8 | 12 | 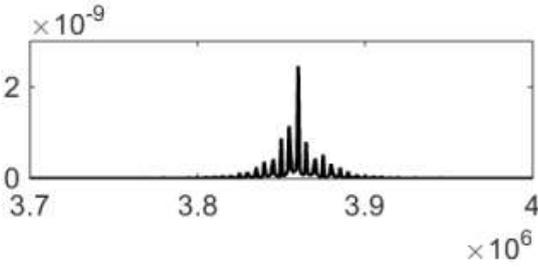 | 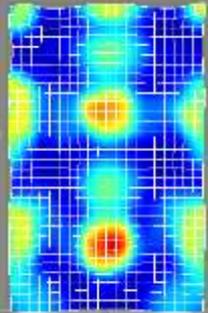 |
| 8 | 14 | 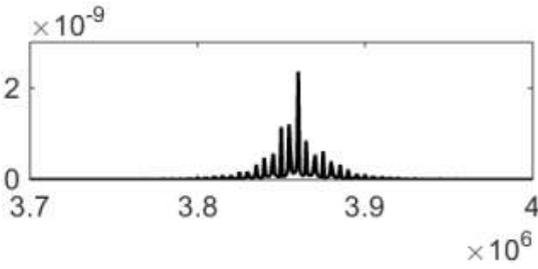 | 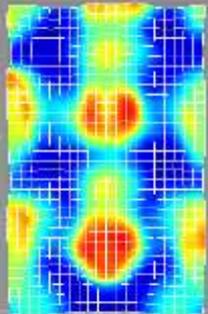 |
| 8 | 16 | 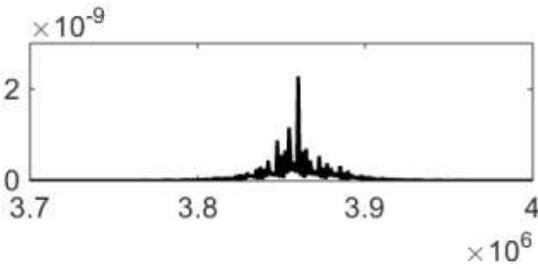 | 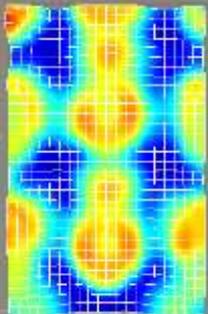 |
| 8 | 18 | 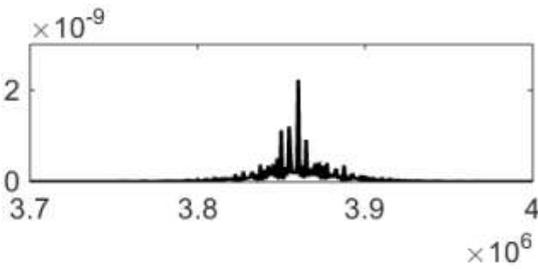 | 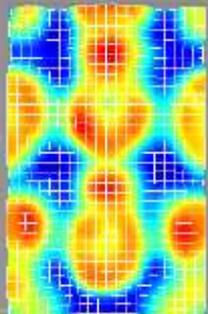 |
| 8 | 20 | 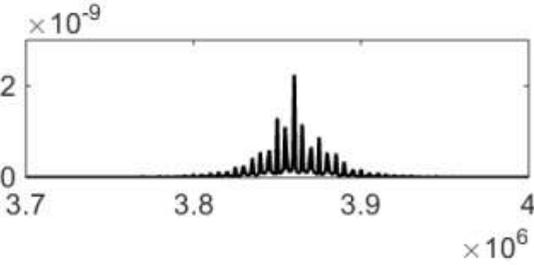 | 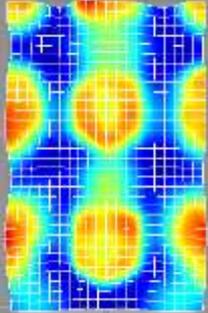 |
| 10 | 0 | 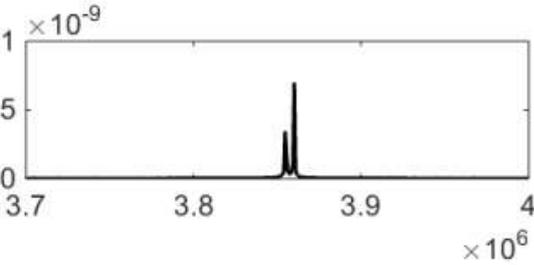 | 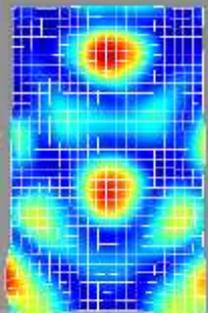 |
| 10 | 2 | 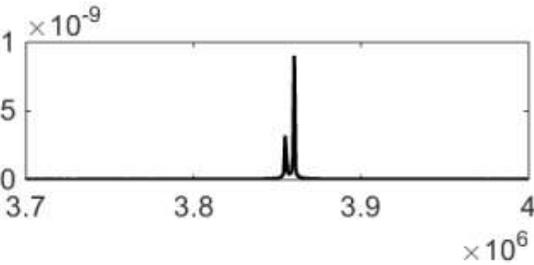 | 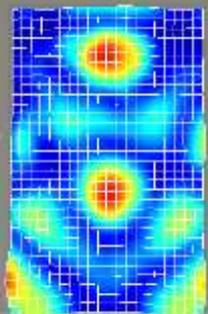 |
| 10 | 4 | 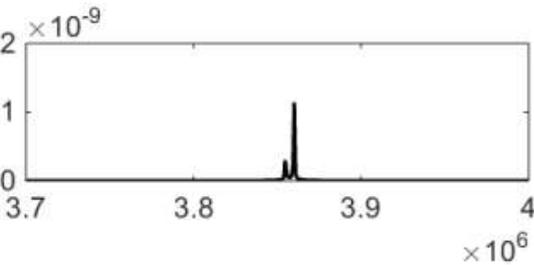 | 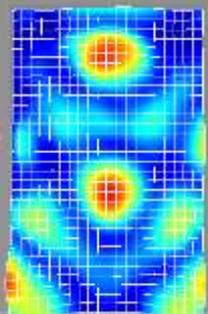 |
| 10 | 6 | 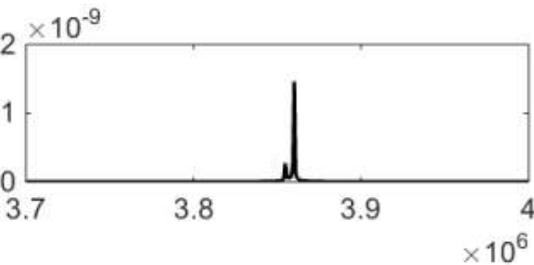 | 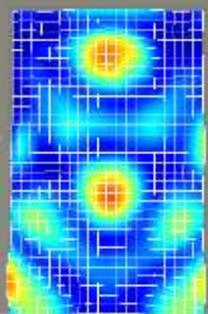 |
| 10 | 8 | 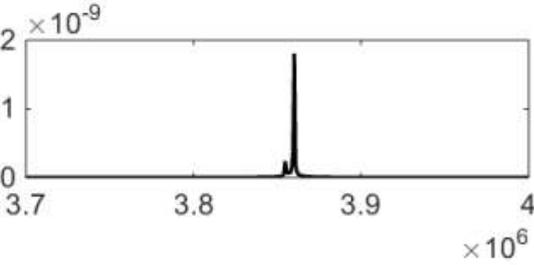 | 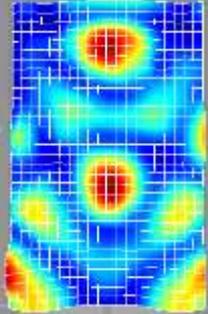 |
| 10 | 10 | 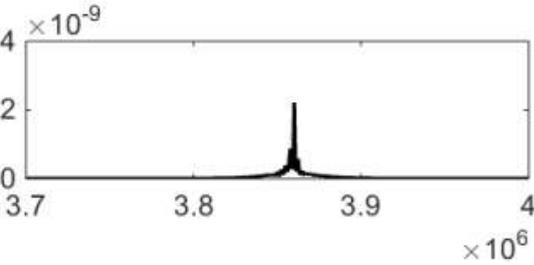 | 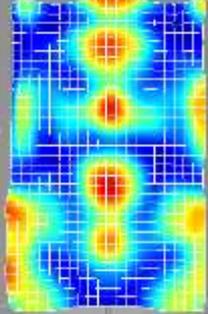 |
| 10 | 12 | 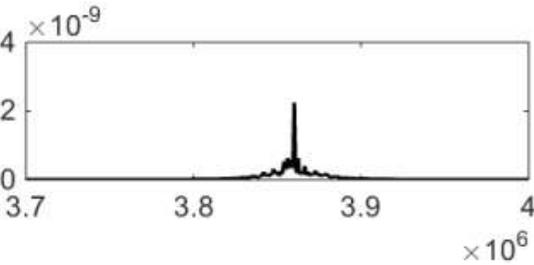 | 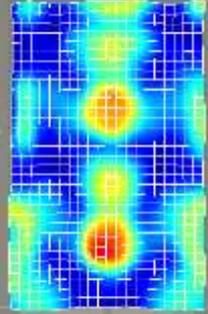 |
| 10 | 14 | 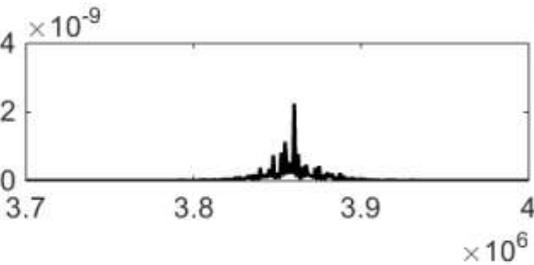 | 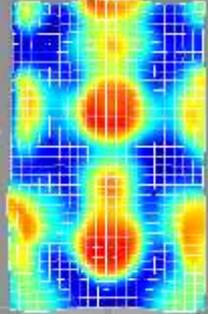 |
| 10 | 16 | 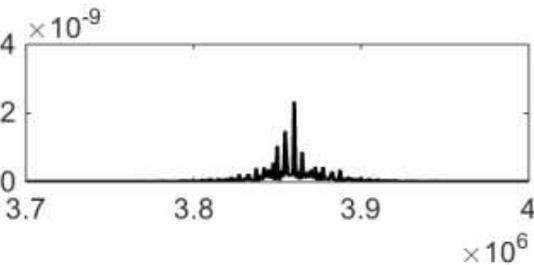 | 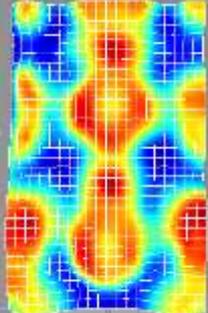 |
| 10 | 18 | 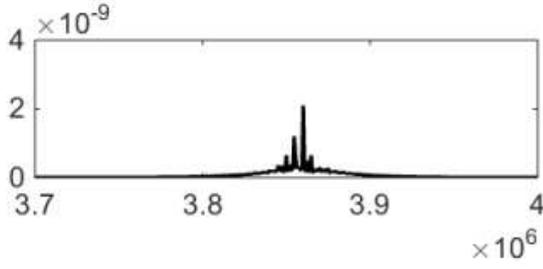 | 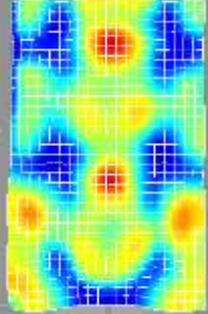 |
| 10 | 20 | 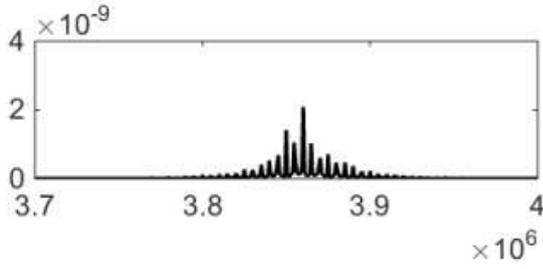 | 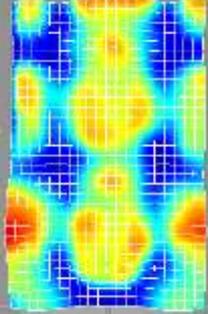 |
| 12 | 0 | 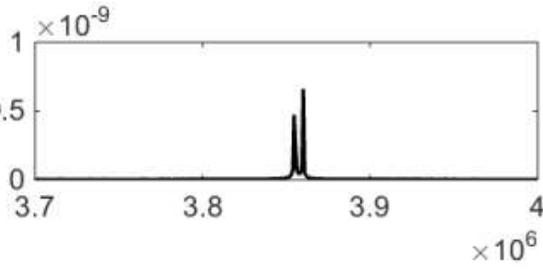 | 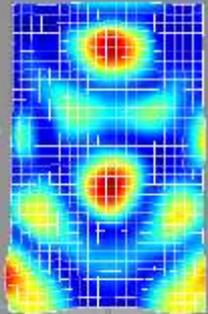 |
| 12 | 2 | 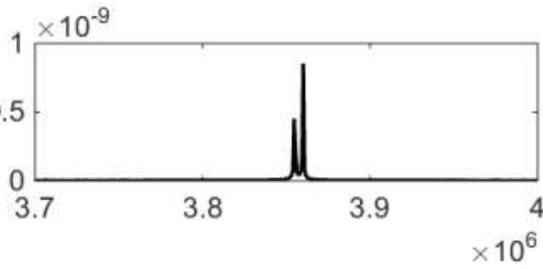 | 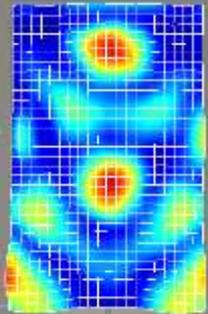 |
| 12 | 4 | 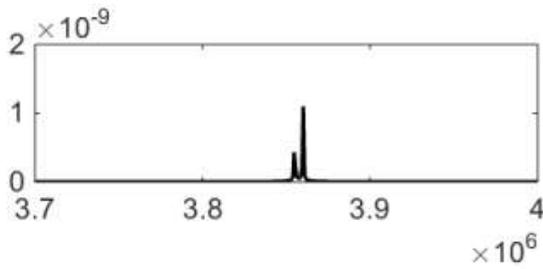 | 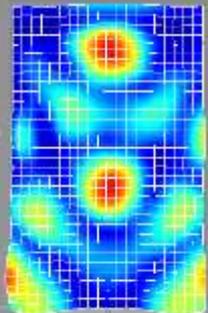 |
| 12 | 6 | 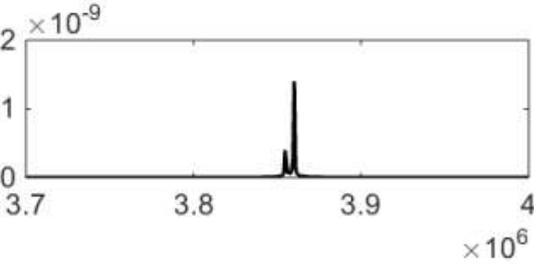 | 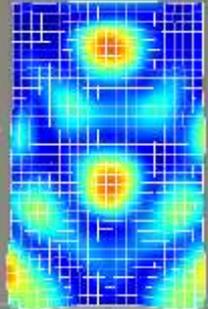 |
| 12 | 8 | 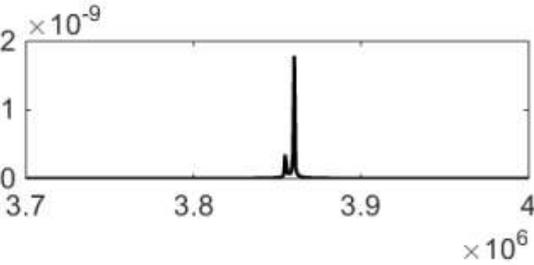 | 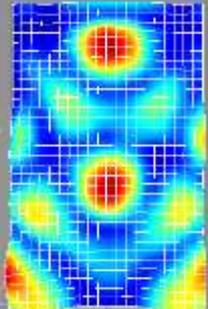 |
| 12 | 10 | 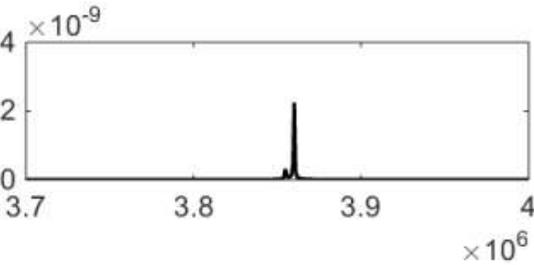 | 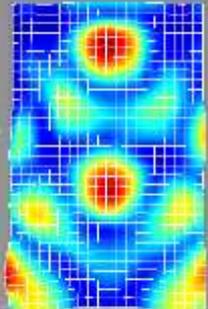 |
| 12 | 12 | 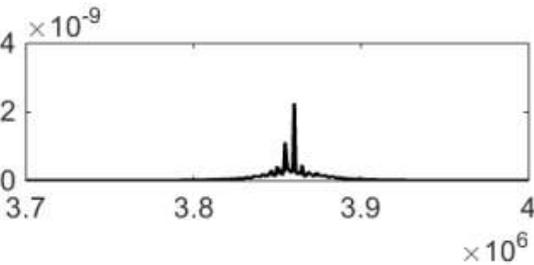 | 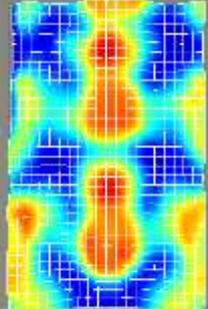 |
| 12 | 14 | 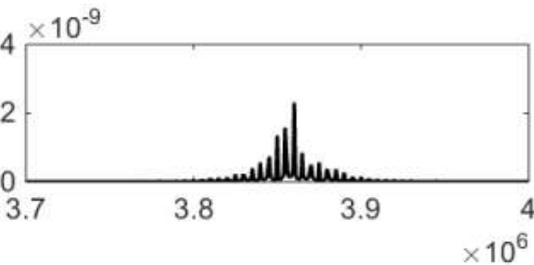 | 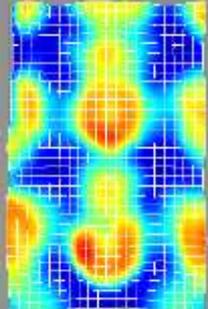 |
| 12 | 16 | 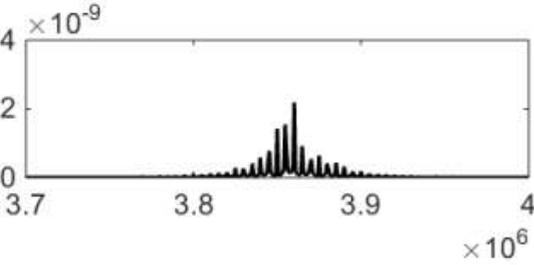 | 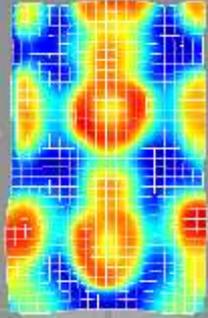 |
| 12 | 18 | 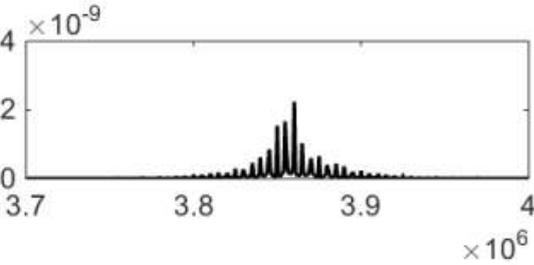 | 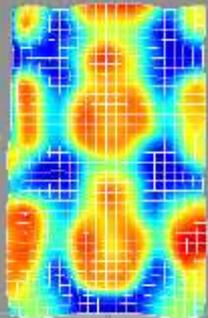 |
| 12 | 20 | 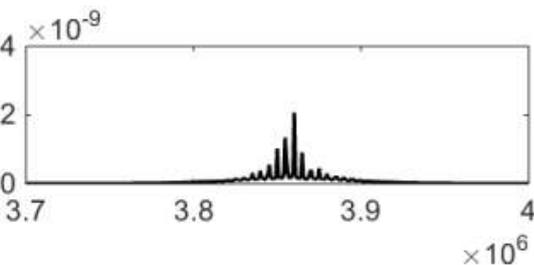 | 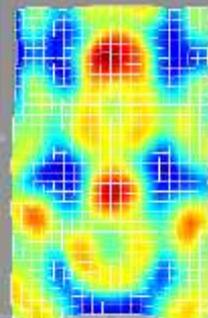 |
| 14 | 0 | 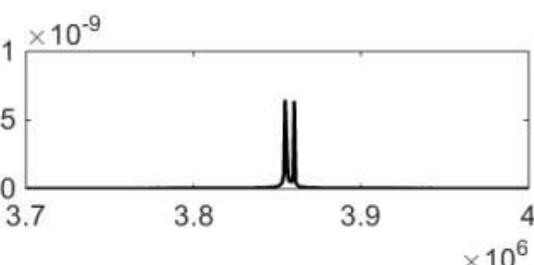 | 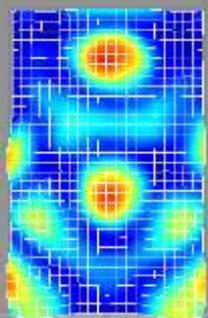 |
| 14 | 2 | 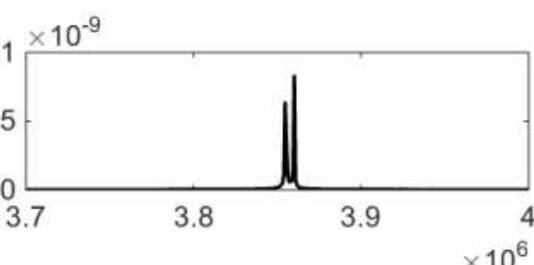 | 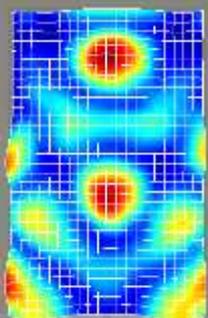 |
| 14 | 4 | 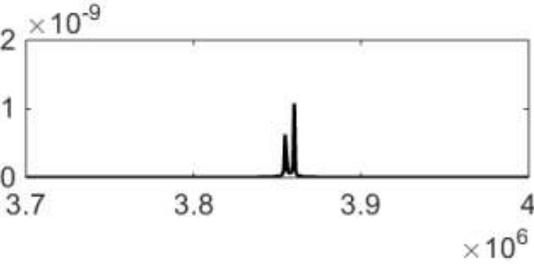 | 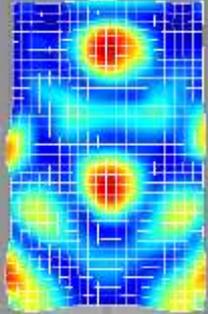 |
| 14 | 6 | 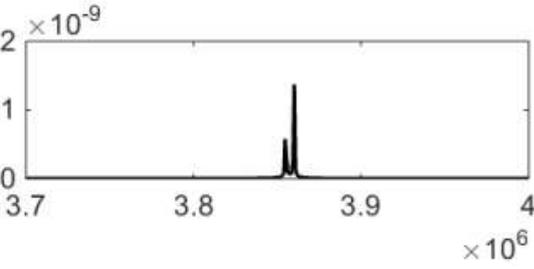 | 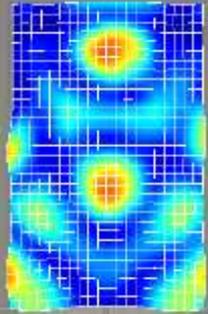 |
| 14 | 8 | 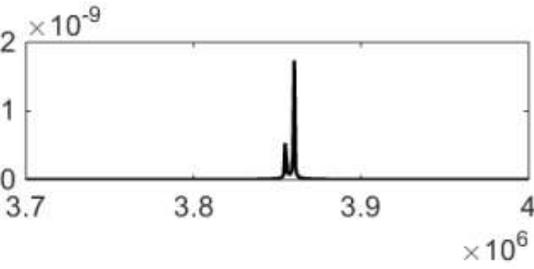 | 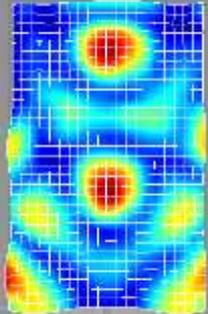 |
| 14 | 10 | 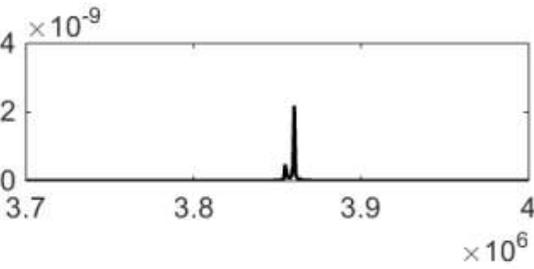 | 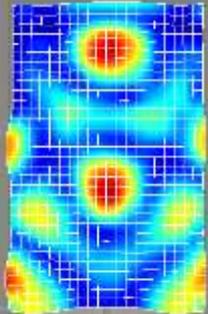 |
| 14 | 12 | 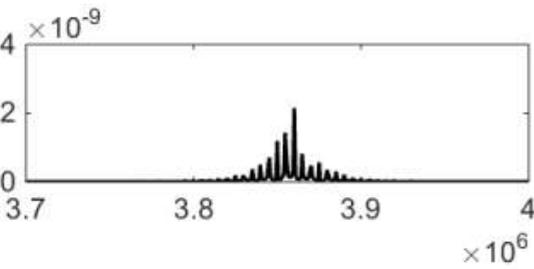 | 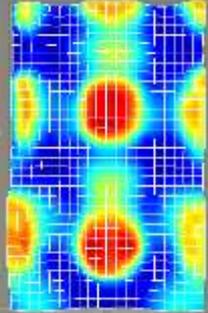 |
| 14 | 14 | 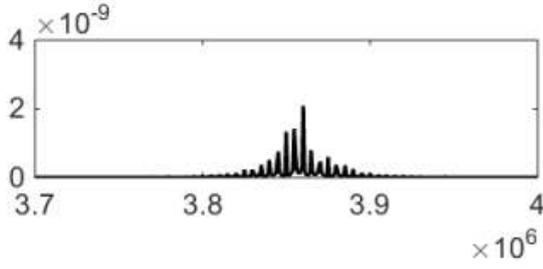 | 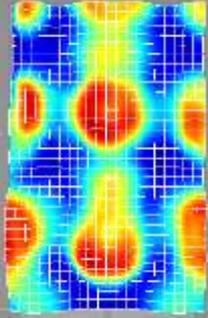 |
| 14 | 16 | 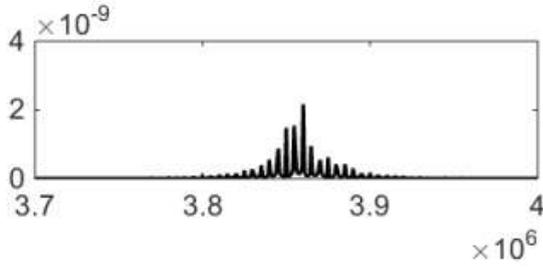 | 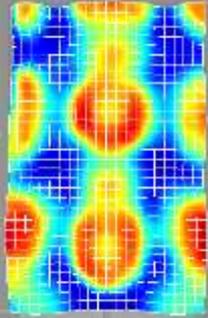 |
| 14 | 18 | 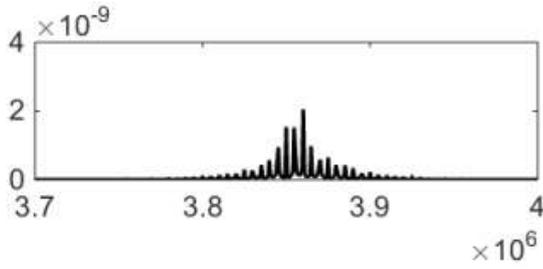 | 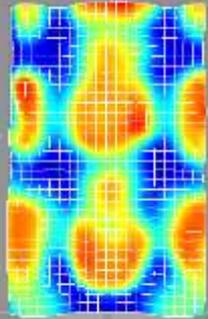 |
| 14 | 20 | 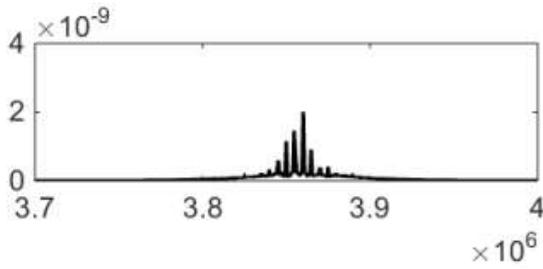 | 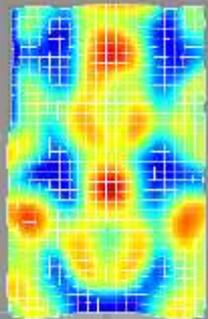 |
| 16 | 0 | 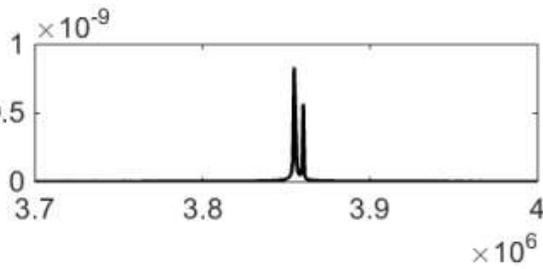 | 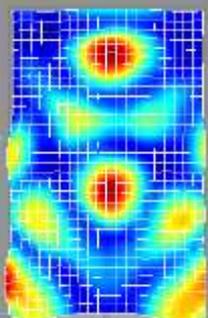 |
| 16 | 2 | 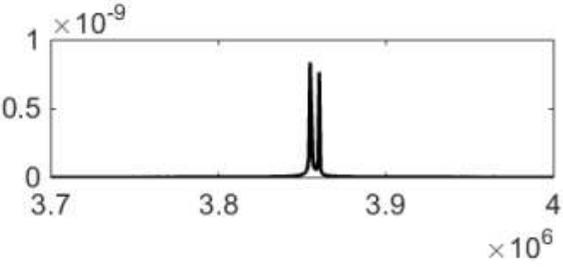 | 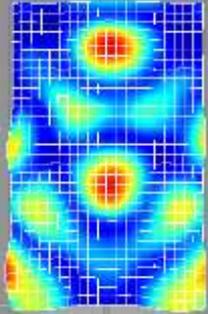 |
| 16 | 4 | 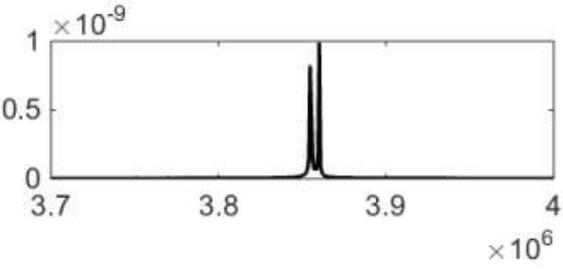 | 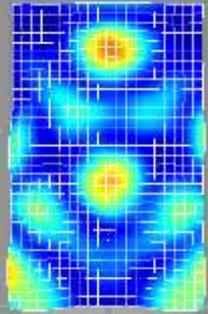 |
| 16 | 6 | 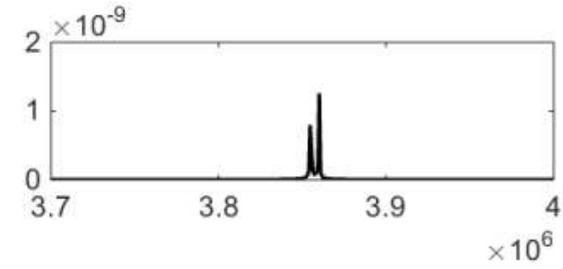 | 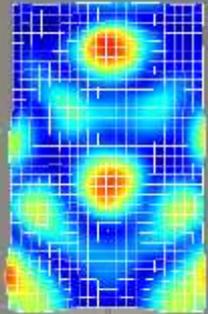 |
| 16 | 8 | 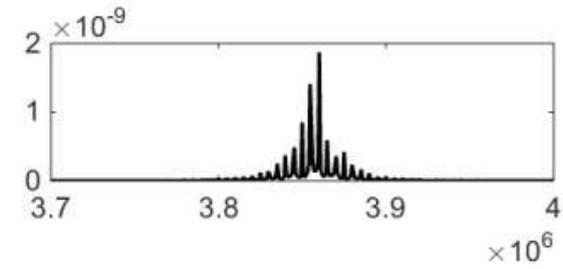 | 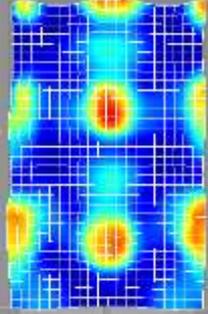 |
| 16 | 10 | 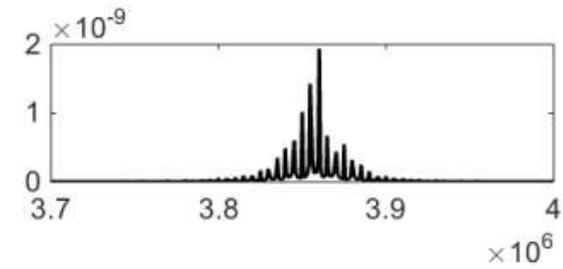 | 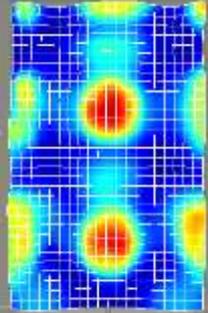 |
| 16 | 12 | 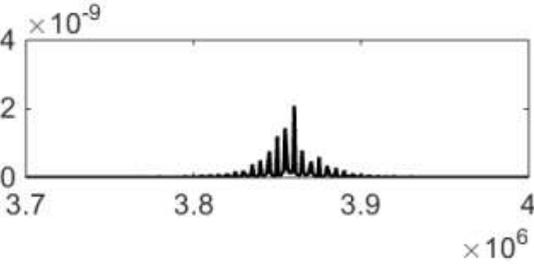 | 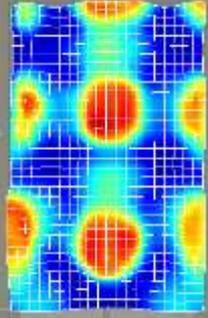 |
| 16 | 14 | 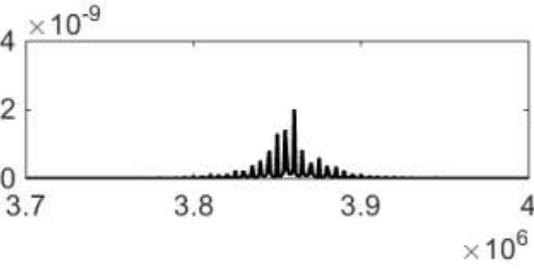 | 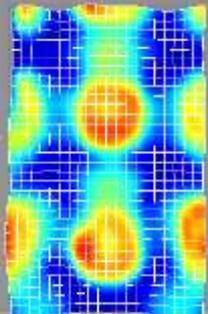 |
| 16 | 16 | 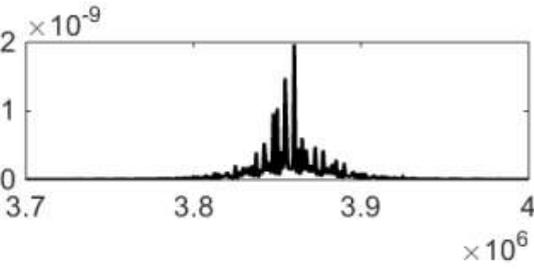 | 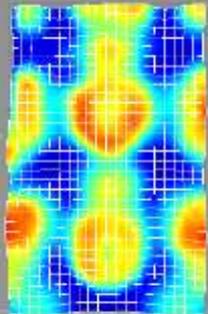 |
| 16 | 18 | 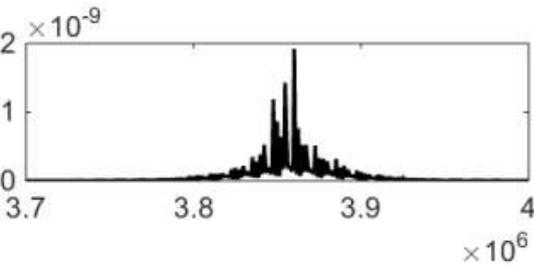 | 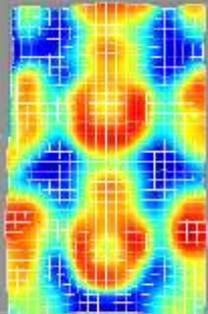 |
| 16 | 20 | 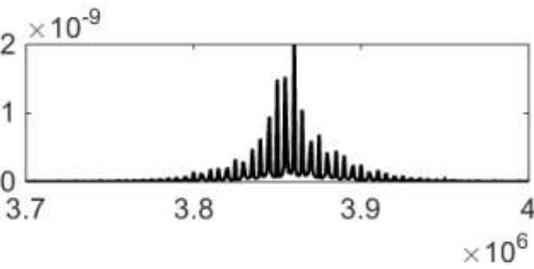 | 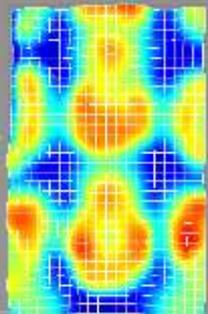 |
| 18 | 0 | 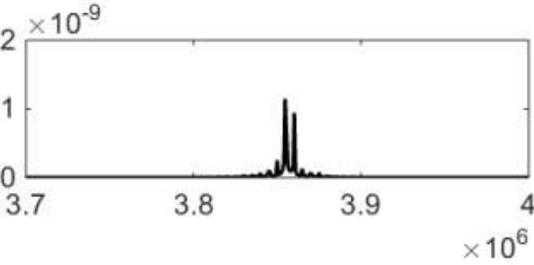 | 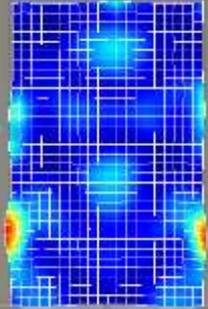 |
| 18 | 2 | 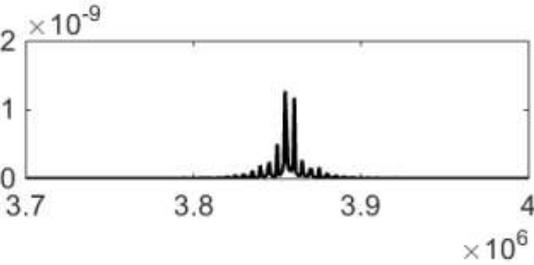 | 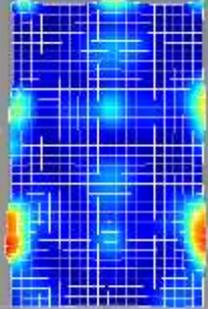 |
| 18 | 4 | 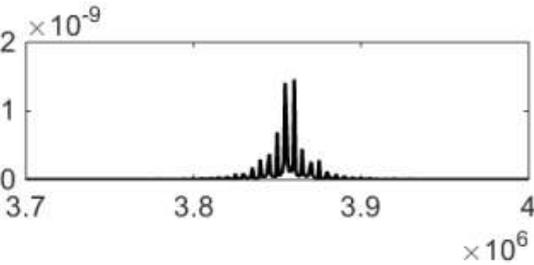 | 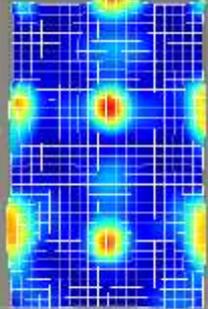 |
| 18 | 6 | 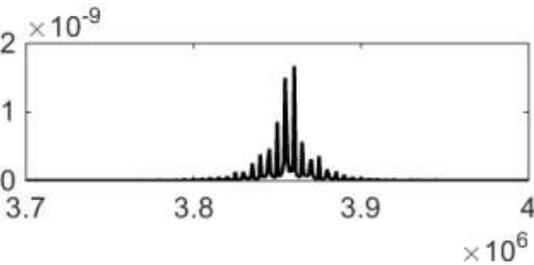 | 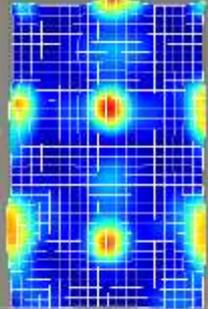 |
| 18 | 8 | 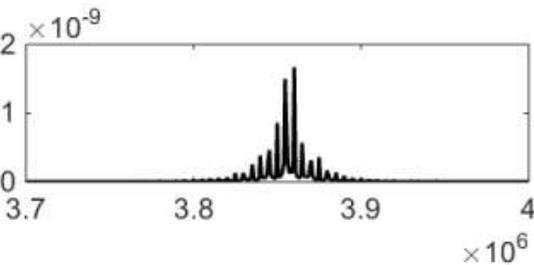 | 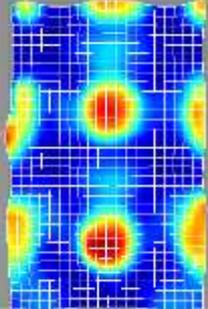 |
| 18 | 10 | 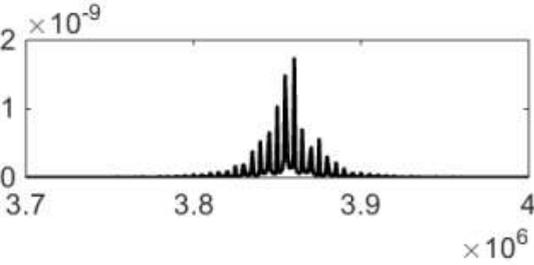 | 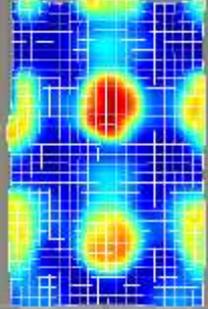 |
| 18 | 12 | 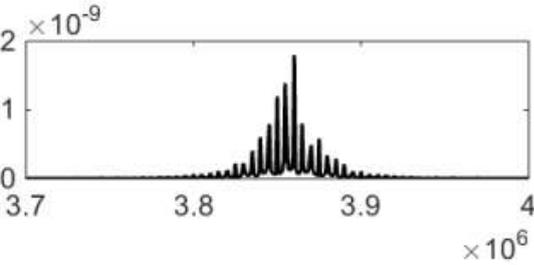 | 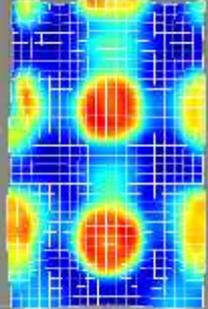 |
| 18 | 14 | 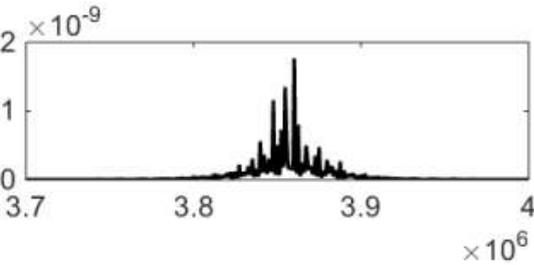 | 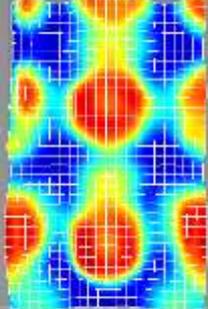 |
| 18 | 16 | 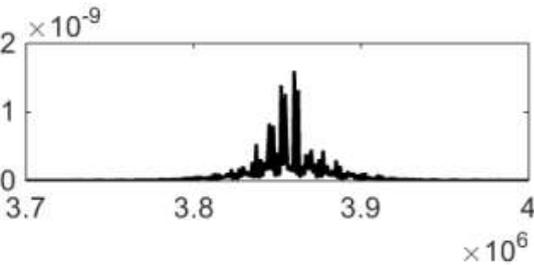 | 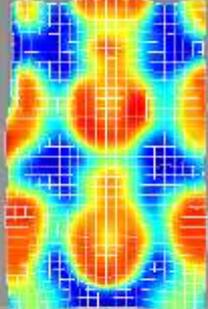 |
| 18 | 18 | 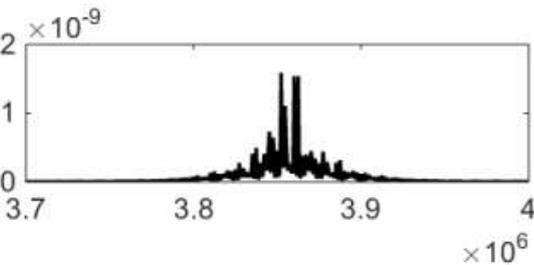 | 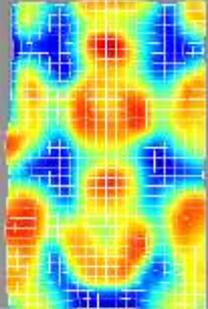 |

| 18 | 20 | 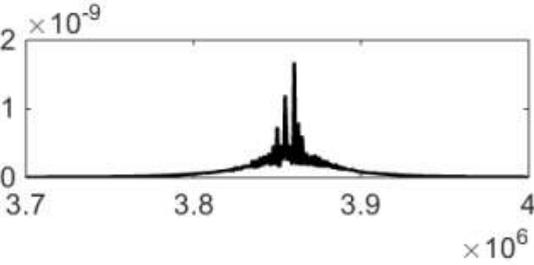 | 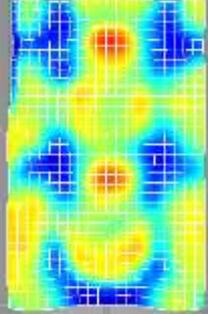 |
|---|---|---|---|
| 20 | 0 | 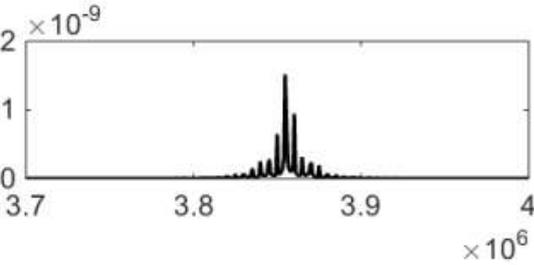 | 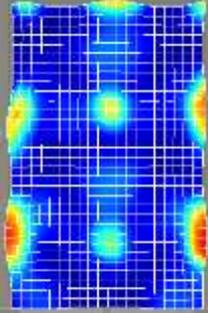 |
| 20 | 2 | 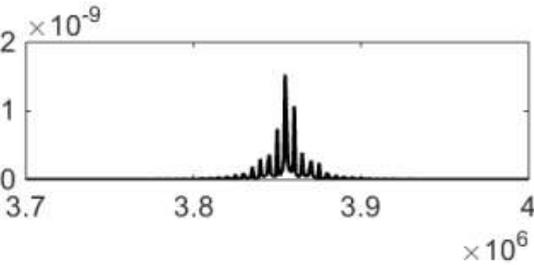 | 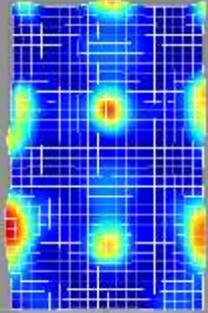 |
| 20 | 4 | 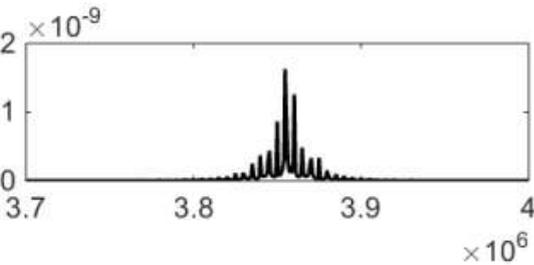 | 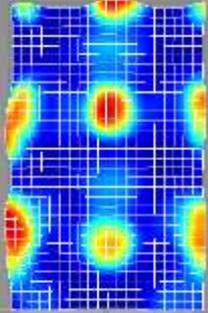 |
| 20 | 6 | 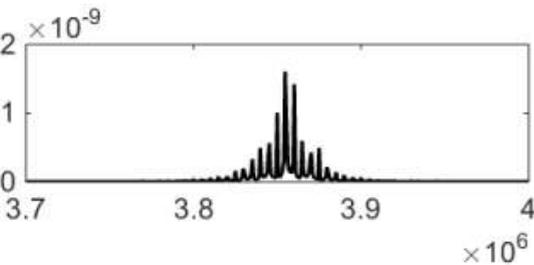 | 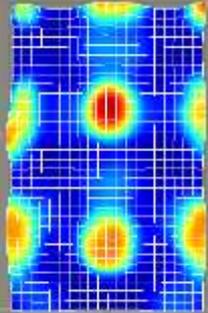 |

| 20 | 8  | 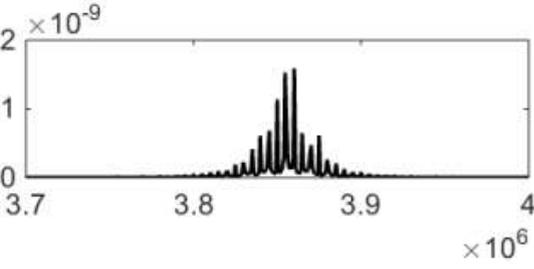 | 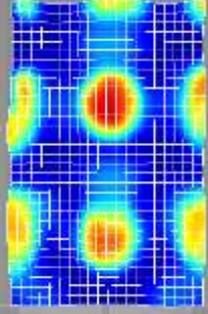 |
|----|----|----------------------|----------------------|
| 20 | 10 | 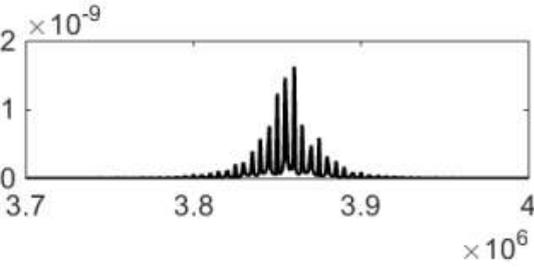 | 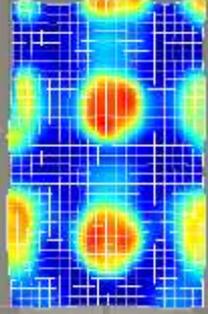 |
| 20 | 12 | 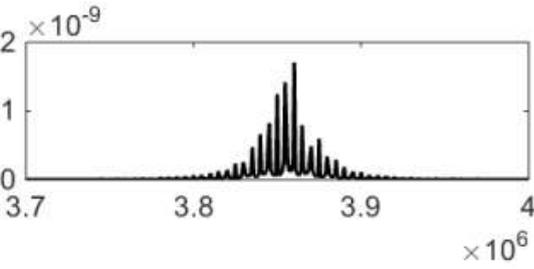 | 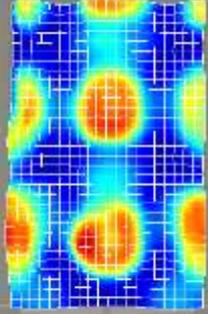 |
| 20 | 14 | 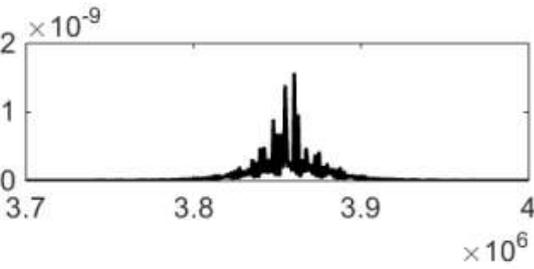 | 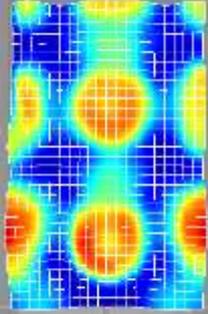 |
| 20 | 16 | 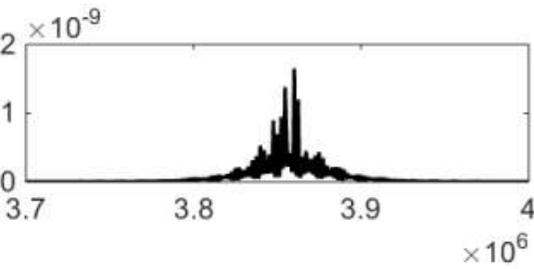 | 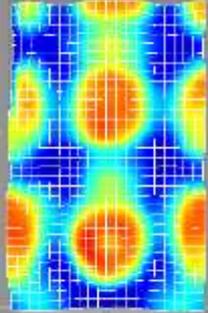 |
| 20 | 18 | 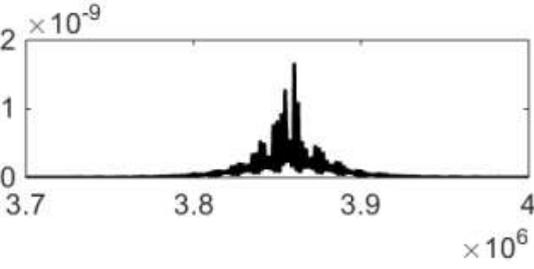 | 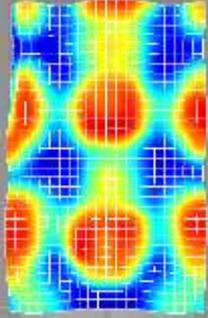 |
| 20 | 20 | 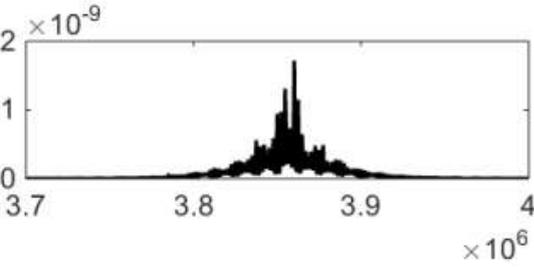 | 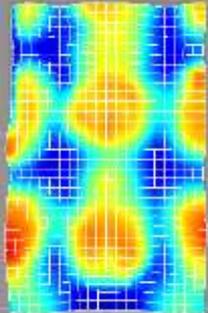 |
| 22 | 0 | 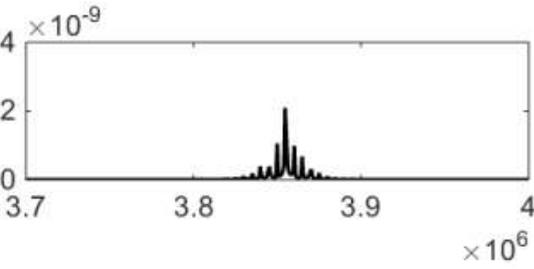 | 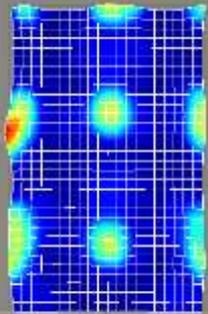 |
| 22 | 2 | 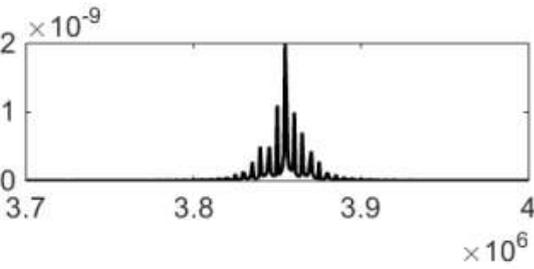 | 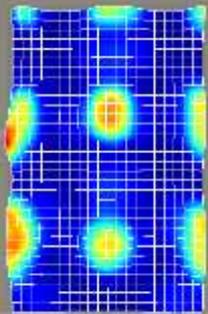 |
| 22 | 4 | 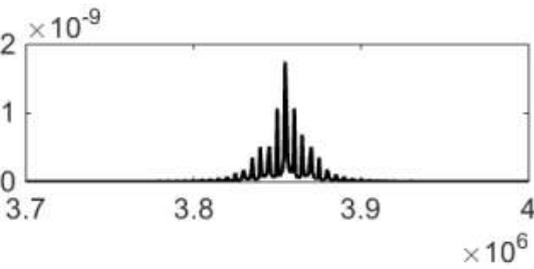 | 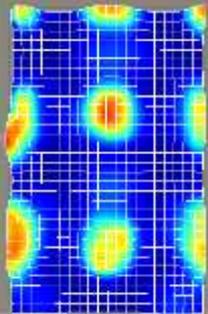 |
| 22 | 6 | 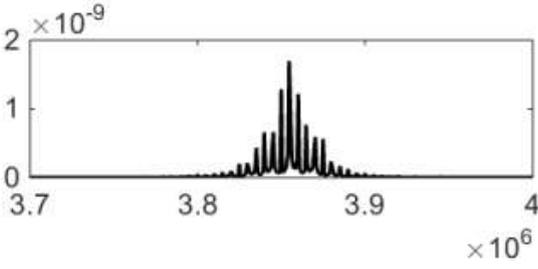 | 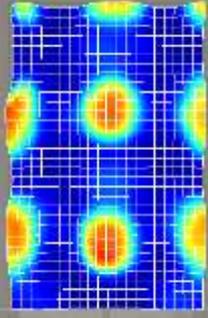 |
| 22 | 8 | 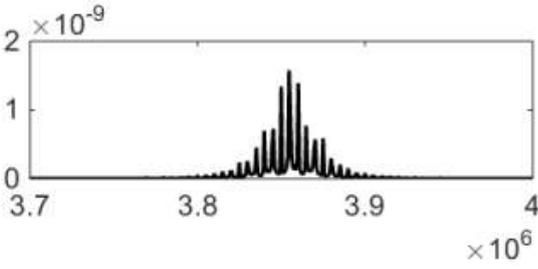 | 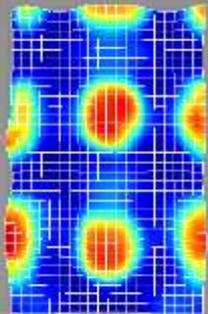 |
| 22 | 10 | 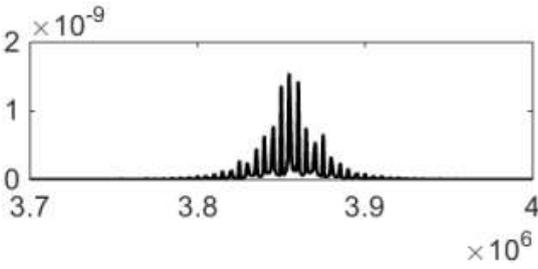 | 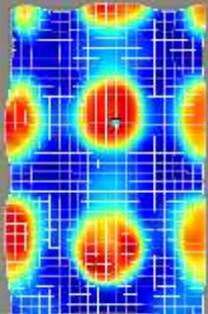 |
| 22 | 12 | 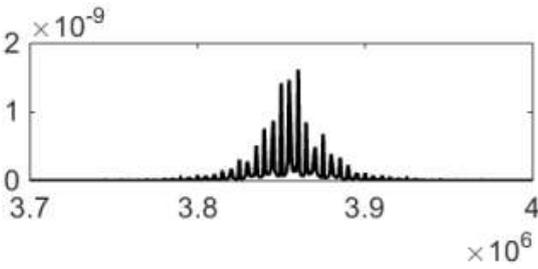 | 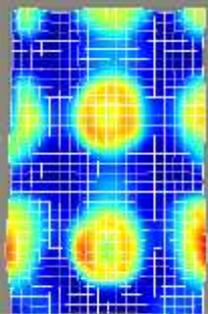 |
| 22 | 14 | 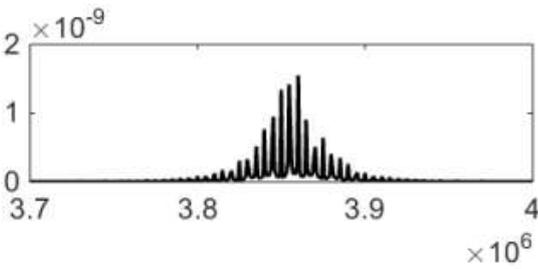 | 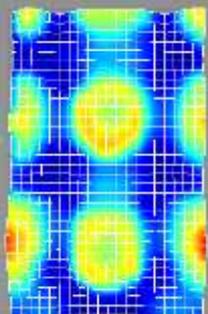 |
| 22 | 16 | 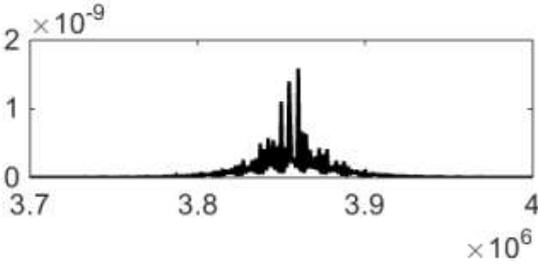 | 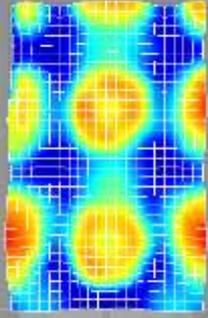 |
| 22 | 18 | 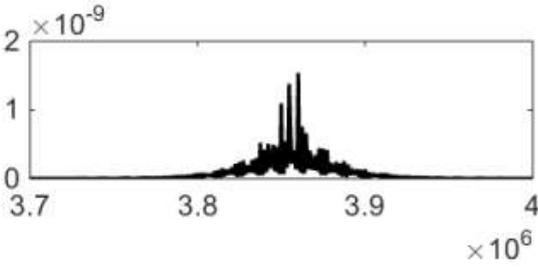 | 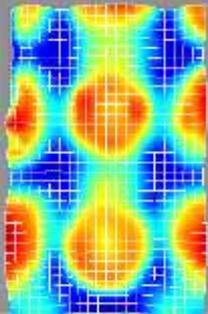 |
| 22 | 20 | 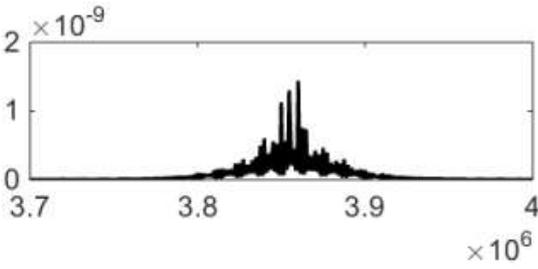 | 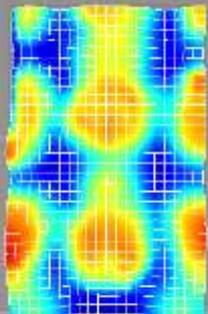 |